\PassOptionsToPackage{pdfpagelabels=false}{hyperref}

\documentclass[fleqn,usenatbib,useAMS]{mnras}

\usepackage{graphicx}
\usepackage{amsmath}
\usepackage[dvipsnames]{xcolor}  
\usepackage{amssymb}
\usepackage{color}
\usepackage{soul}
\usepackage{ulem}
\usepackage[figure,figure*,table,table*]{hypcap}

\input{macros.sty}

\title[BOSS lensing discrepancy]
{New perspectives on the BOSS small-scale lensing discrepancy for the Planck $\Lambda$CDM Cosmology}
\author[J.~U.~Lange et al.]
{Johannes~U.~Lange$^1$\thanks{email: johannesulf.lange@yale.edu}, Xiaohu Yang$^{2, 3}$, Hong Guo$^4$, Wentao Luo$^{2,5}$ and \newauthor Frank~C.~van~den~Bosch$^1$\\
	$^1$Department of Astronomy, Yale University, P.O. Box 208101, New Haven, CT 06511, USA\\
	$^2$Department of Astronomy, School of Physics and Astronomy, Shanghai Jiao Tong University, Shanghai 200240,  China\\
	$^3$Tsung-Dao Lee Institute, and Shanghai Key Laboratory for Particle Physics and Cosmology, Shanghai Jiao Tong University, \\ Shanghai 200240,  China\\
	$^4$Key Laboratory for Research in Galaxies and Cosmology, Shanghai Astronomical Observatory, Shanghai 200030, China\\
	$^5$Kavli Institute for the Physics and Mathematics of the Universe (WPI), University of Tokyo, Kashiwa 277-8583, Japan}

\pubyear{2018}

\begin{document}
	
	\date{Accepted xxx. Received xxx}
	
	\label{firstpage}
	\pagerange{\pageref{firstpage}--\pageref{lastpage}}
	
	\maketitle
	
	\begin{abstract}
		We investigate the abundance, small-scale clustering and galaxy-galaxy lensing signal of galaxies in the \textit{Baryon Oscillation Spectroscopic Survey} (BOSS). To this end, we present new measurements of the redshift and stellar mass dependence of the lensing properties of the galaxy sample. We analyse to what extent models assuming the Planck18 cosmology fit to the number density and clustering can accurately predict the small-scale lensing signal. In qualitative agreement with previous BOSS studies at redshift $z \sim 0.5$ and with results from the \textit{Sloan Digital Sky Survey}, we find that the expected signal at small scales ($0.1 < r_\rmp < 3 \Mpch$) is higher by $\sim 25\%$ than what is measured. Here, we show that this result is persistent over the redshift range $0.1 < z < 0.7$ and for galaxies of different stellar masses. If interpreted as evidence for cosmological parameters different from the Planck CMB findings, our results imply $S_8 = \sigma_8 \sqrt{\Omega_\rmm / 0.3} = 0.744 \pm 0.015$, whereas $S_8 = 0.832 \pm 0.013$ for Planck18. However, in addition to being in tension with CMB results, such a change in cosmology alone does not accurately predict the lensing amplitude at larger scales. Instead, other often neglected systematics like baryonic feedback or assembly bias are likely contributing to the small-scale lensing discrepancy. We show that either effect alone, though, is unlikely to completely resolve the tension. Ultimately, a combination of the two effects in combination with a moderate change in cosmological parameters might be needed.
	\end{abstract}
	
	
	\section{Introduction}
	\label{sec:intro}
    
	The spatial distribution of galaxies and matter in the Universe encodes important constraints on the concordance $\Lambda$ $+$ Cold Dark Matter ($\Lambda$CDM) cosmological model. We are entering an era in which the cosmological information extracted from large-scale structure surveys meets or exceeds the information content of the cosmic microwave background (CMB). Cosmological constraints from large-scale structure surveys can be inferred, for example, from a combination of galaxy clustering and galaxy-galaxy lensing \citep[e.g.][]{Cacciato_13, Mandelbaum_13, Leauthaud_17, DES_18a, Singh_18}, cosmic shear \citep[e.g.][]{Fu_14, Jee_16, Hildebrandt_17, Hikage_19, Chang_19}, redshift-space distortions \citep[e.g.][]{Yang_04, Reid_14, Zhai_19} and the abundance and properties of galaxy clusters \citep[e.g.][]{Rykoff_14, Costanzi_18}.
	
	Galaxy clustering and galaxy-galaxy lensing are measurements of the galaxy-galaxy and the galaxy-matter correlation functions, respectively. The combined analysis of both quantities allows to break degeneracies regarding the galaxy bias and to indirectly infer the matter-matter correlation function, thereby probing cosmology. Intriguingly, several recent studies \citep[see e.g.][]{Cacciato_13, Mandelbaum_13, Leauthaud_17, DES_18a, Singh_18} modelling these two observables have reported various levels of tension in cosmological parameters with respect to the \citet[][hereafter Planck18]{Planck_18} CMB analysis. Other large-scale structure studies, particularly those using cosmic shear, have reported similar tensions \citep[see e.g.][]{Reid_14, Fu_14, Jee_16, Hildebrandt_17, Hikage_19}. Of particular importance is the combined constraint on $\Omega_\rmm$, the matter density of the Universe, and $\sigma_8$, a measurement of the strength of matter fluctuations. Generally, large-scale structure studies prefer relatively low values for $\Omega_\rmm$ and $\sigma_8$ compared to Planck18. This discrepancy could, for example, hint at new physics beyond the standard $\Lambda$CDM model \citep{More_13b, Leauthaud_17}.
	
	On large, quasi-linear scales, the matter and galaxy distributions can be calculated analytically and are insensitive to details of galaxy formation physics. On the other hand, smaller, non-linear scales hold larger statistical constraining power but are also more difficult to model. The focus of this work will be the latter. Small scales are particularly sensitive to many details of the relationship between galaxies and the dark matter haloes that host them, known as the galaxy-halo connection \citep[e.g.][]{Berlind_02, Cacciato_12}. Of particular concern is the issue of assembly bias, the fact that the clustering of dark matter haloes depends on halo properties other than mass \citep{Gao_05, Wechsler_06, Villarreal_17, Salcedo_18}, and its manifestation in galaxy populations \citep[see e.g.][]{Zentner_14, Zentner_19, More_16}, which we will call galaxy assembly bias. Additionally, baryonic physics, particularly in the form of powerful energetic feedback from active galaxy nuclei (AGN), might be able to affect the matter distribution on such scales \citep[e.g.,][]{Zentner_13, Springel_18}. Both these effects will be somewhat degenerate with changes in cosmological parameters.

    The goal of this paper is to investigate to which extent these three competing effects, cosmological parameters, galaxy assembly bias and baryonic physics, can reconcile the observed clustering and lensing of galaxies in the Baryon Oscillation Spectroscopic Survey (BOSS) with the $\Lambda$CDM prediction. We will first present new measurements for the clustering and lensing properties of BOSS galaxies as a function of redshift and stellar mass. We then show that under Planck18 cosmological parameters and no baryonic physics and galaxy assembly bias, the small-scale lensing signal is incompatible with the clustering strength, independent of stellar mass or redshift. We then investigate to what extent the three aforementioned effects can explain both the magnitude of the lensing discrepancy, as well as its dependence on the galaxy sample in question.
	
	The paper is organised as follows. In Section \ref{sec:data}, we describe our measurements of the abundance, clustering and lensing of BOSS galaxies. We describe in Section \ref{sec:model} our model for predicting those quantities in a $\Lambda$CDM cosmology and a model for the galaxy-halo connection. We compare our lensing predictions to observations in Section \ref{sec:discrepancy}. In Section \ref{sec:explaining}, we investigate to what extent changes in cosmological parameters, galaxy assembly bias and baryonic feedback can reconcile the predictions with observational results. We discuss our findings in the context of other cosmological studies in Section \ref{sec:cosmology_comparison} and present our conclusion in Section \ref{sec:conclusion}. Unless otherwise noted, we assume the best-fit Planck18 (TT,TE,EE+lowE+lensing) cosmological parameters, $\Omega_\rmm = 0.3153$, $\sigma_8 = 0.8111$ $h = H_0 / 100 \kms = 0.6736$, $n_s = 0.9649$ and $\Omega_\rmb = 0.0493$, and neglect the effect of massive neutrinos. Finally, throughout this paper, $\log$ denotes the logarithm with respect to base $10$.
	
	\section{Data}
	\label{sec:data}
	
	In this paper, we use the same galaxy catalogues as in \cite{Guo_18} from data release 12 of the SDSS-III BOSS \citep{Reid_16}, which covers a sky area of $\sim 10,000 \ \mathrm{deg}^2$ and a broad redshift range of $0.1<z<0.8$. The BOSS sample is composed of two subsamples targeting luminous galaxies at low and high redshifts, denoted as the LOWZ and CMASS samples, respectively. The detailed target selection cuts can be found in \cite{Eisenstein_11} and \cite{Dawson_13}. We use the whole BOSS galaxy sample in the redshift range of $0.1 < z < 0.7$ to measure the clustering and lensing signals. 

    \begin{figure}
        \centering
        \includegraphics[width=\columnwidth]{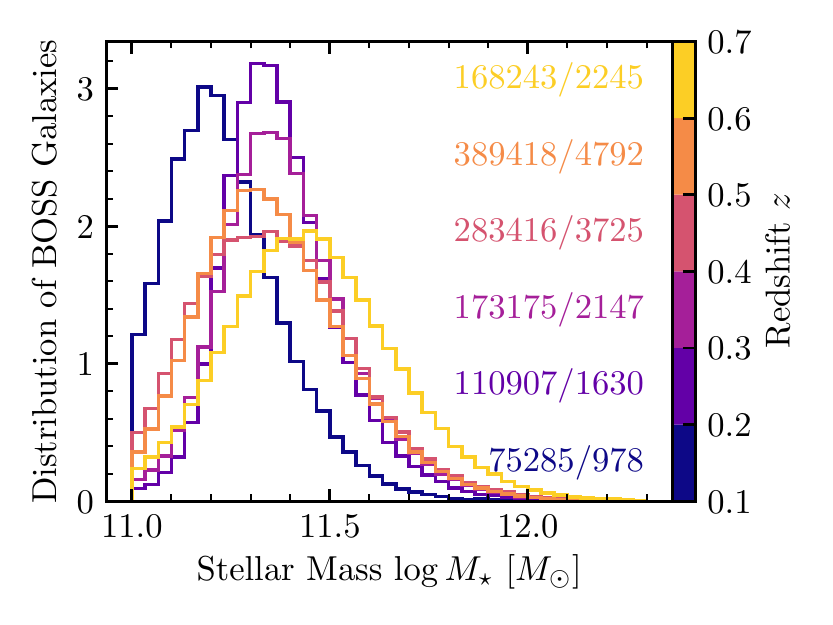}
        \caption{The stellar mass and redshift distribution of BOSS galaxies used in this study. The numbers on the right denote the number of BOSS galaxies used to determine the clustering strength (left) and the galaxy-galaxy lensing signal (right). The significantly lower number of galaxies available for lensing is due to the small overlap of the BOSS and the CFHTLenS footprints.}
		\label{fig:gal_distribution}
	\end{figure}

	For each galaxy in the sample, a stellar mass has been estimated in \cite{Chen_12} by spectral fitting with the principal component analysis (PCA) method. We use the same stellar mass estimates as in \cite{Guo_18}, with the initial mass function (IMF) of \cite{Chabrier_03}, the stellar population synthesis (SPS) model of \cite{Bruzual_03} and the time-dependent dust attenuation law of \cite{Charlot_00}. It is worth noting that stellar mass estimates can very substantially depending on the code used and intrinsic assumptions \citep{Bundy_15, Bundy_17, Tinker_17}. However, we do not expect stellar mass changes to impact the conclusions of this paper since we marginalise over the galaxy-halo connection and are primarily concerned with cosmology. As the BOSS sample targets luminous galaxies at different redshifts, the majority of the BOSS galaxies have a stellar mass, $M_\star$, larger than $10^{11}\Msun$ and we select only those galaxies for our subsequent analysis. We refer the readers to \cite{Guo_18} for more details of the galaxy sample. Our total sample consists of $\sim 1.2$ million galaxies with $M_\star \geq 10^{11} \Msun$ and $0.1 < z < 0.7$. The stellar mass and redshift distribution as well as the number of galaxies used to determine the clustering and lensing signal is shown in Fig.~\ref{fig:gal_distribution}. In the absence of incompleteness, the galaxy numbers would rise towards lower stellar masses. Therefore, the decreasing number of galaxies at low stellar masses indicates high amount of incompleteness for lower-mass galaxies. Throughout this work, we always use the incomplete stellar mass functions of \cite{Guo_18} when modelling galaxy abundances.

	\subsection{Clustering}
    \label{sec:clustering}
    
	We measure the projected two-point correlation function (2PCF) of the BOSS galaxies using the same methodology as in \cite{Guo_18}. Specifically, we use the Landy--Szalay estimator \citep{Landy_93} to estimate the anisotropic 2PCF $\xi(r_\rmp, r_\pi)$,
	\begin{equation}
	    \xi(r_\rmp, r_\pi) = \frac{\mathrm{DD} - 2\mathrm{DR} + \mathrm{RR}}{\mathrm{RR}},
	\end{equation}
	where $r_\rmp$ and $r_\pi$ are the comoving separations perpendicular and along the line of sight (LOS), respectively. $\mathrm{DD}$, $\mathrm{DR}$ and $\mathrm{RR}$ are the normalised numbers of data-data, data-random and random-random pairs for these separations, respectively. To reduce the shot noise errors, we use 2 million random points from the random catalogue as detailed in \cite{Reid_16}. The projected 2PCF is obtained by integrating along the LOS,
	\begin{equation}
	    w_\rmp (r_\rmp) = \int\limits_{-r_{\pi, \rm max}}^{r_{\pi, \rm max}} \xi(r_\rmp, r_\pi) \rmd r_\pi.
	\end{equation}
    We choose $r_{\pi, \rm max} = 100 \Mpch$ as the integration distance.
	
	The projected 2PCF is measured separately for six redshift intervals between $z = 0.1$ and $z = 0.7$ with a bin size of $\Delta z = 0.1$. We measure $w_\rmp (r_\rmp)$ for all galaxies with stellar masses above $10^{11} \Msun$, and for two subsamples with  $\log(M_\star / \Msun)$ in the ranges $[11.0, 11.5]$ and $(11.5, 12.0]$, respectively. The LOS separation $r_{\rm p}$ is binned with a logarithmic width of $\Delta\log r_{\rm p} = 0.2$ from $0.63$ to $63\Mpch$, while $r_\pi$ is in linear bins of $\Delta r_\pi = 2\Mpch$ from 0 to $100\Mpch$. The range for $r_\rmp$ is motivated by fibre collisions at the lower end and by a decreasing signal-to-noise ratio for large $r_\rmp$. The measurement errors are estimated from the jackknife resample technique with 100 subsamples \citep{Guo_13a,Guo_14a}. In the case that $w_\rmp (r_\rmp)$ is measured for two galaxy samples with different stellar masses, we compute their respective covariance matrices, which we include as measurement errors in our analysis. 
    
    Our clustering measurements for all galaxies above $10^{11} \, \Msun$ in stellar mass are shown in the upper panel of Fig.~\ref{fig:hod_fits}. We find the clustering amplitude to be roughly independent of redshift, in agreement with \cite{Saito_16} and \cite{Guo_18}. The clustering measurements for the two subsamples split by stellar mass are not shown but are almost exactly the ones used in \cite{Guo_18}\footnote{Compared to \cite{Guo_18} we use more randoms.}. For example, we find that the clustering strength of the high-stellar mass sample is consistently higher by a factor of $\sim 2$, independent of $r_\rmp$ and $z$. This indicates that more massive galaxies live on average in more massive haloes.
    
	\subsection{Galaxy-galaxy lensing}
    \label{sec:gglens}
    
	We use the publicly available shape catalog\footnote{\url{http://www.cadc-ccda.hia-iha.nrc-cnrc.gc.ca/en/community/CFHTLens/query.html}} of the \textit{Canada-France-Hawaii Telescope Lensing Survey} (CFHTLenS) to measure the galaxy-galaxy lensing signal. We closely follow the procedure outlined in \cite{Leauthaud_17}. From the CFHTLenS catalogue we select all sources with positive weight ({\sc weight > 0}), a galaxy classification ({\sc fitclass = 0}), an i-band magnitude $i \leq 24.7$ ({\sc MAG\_i <=24.7}) and {\sc MASK <= 1}. We discard field W2 due to the small overlap with BOSS. We apply a correction to the $e_2$ component of all ellipticities, as discussed in \cite{Heymans_12}. As lenses, we select all BOSS galaxies lying inside the CFHTLenS survey area. For each lens-source pair we require $z_\rms > z_\rml + 0.1$ and $z_\rms > z_\rml + \sigma_{95, \rms} / 2.0$, where $z_\rms$ is the best-fitting photometric redshift of the source, $\sigma_{95, \rms}$ the $95\%$ range of the photometric redshift posterior and $z_\rml$ the spectroscopic redshift of the lens \citep{Leauthaud_17}.
	
	The gravitational potential around the lens galaxy will cause shear distortions of the shape of the background galaxies. In the weak lensing regime, this will induce a tangential shear component $\gamma_\rmt$,
	\begin{equation}
	    \gamma_\rmt (r_\rmp) = \frac{\Delta\Sigma (r_\rmp)}{\Sigma_{\rm crit}}.
	\end{equation}
	The quantity we aim to measure is the so-called excess surface density (ESD),
	\begin{equation}
	    \Delta\Sigma (r_\rmp) = \langle \Sigma (< r_\rmp) \rangle - \Sigma (r_\rmp),
	\end{equation}
	with $\Sigma$ being the projected matter density along the LOS and $\langle \Sigma (< r_\rmp) \rangle$ the average surface density inside the projected radius $r_\rmp$. Finally, the critical surface density is given by
	\begin{equation}
		\Sigma_\mathrm{crit} = \frac{c^2}{4\pi G}\frac{D_\rms}{D_\rml D_{\rm ls} (1 + z_\rml)^2},
	\end{equation}
	where $D_\rms$, $D_\rml$ and $D_{\rm ls}$ are the observer-source, observer-lens and lens-source angular diameter distances, respectively.
	
	The tangential ellipticity of the source galaxy, $e_\rmt$, is related to the ellipticity components $e_1$ and $e_2$ in CFHTLenS via
	\begin{equation}
        e_\rmt = e_1 (2 \sin^2\phi - 1) - 2 e_2 \sin\phi \cos\phi,
	\end{equation}
	where $\phi$ is the angle between the right ascension direction and the line connecting source and lens \citep{Miyatake_15}. The shape of an individual source galaxy is dominated by its intrinsic shape, i.e. it has a non-zero tangential ellipticity $e_{\rm int} \neq 0$. In practice, one needs to average over a substantial number of lens-source pairs and their tangential ellipticities $e_\rmt = \gamma_\rmt + e_{\rm int}$ to get a reliable estimate of the ESD. We thus use as an estimator
	\begin{equation}
		\Delta\Sigma (r_\rmp) = \frac{\sum_{i = 1}^{N_\rml} w_i^{\rml} \sum_{j=1}^{N_{\rms, i}} w_{ij}^{\rm ls} e_{\rmt, ij} \Sigma_{\mathrm{crit}, ij}}{\sum_{i = 1}^{N_\rml} w_i^\rml \sum_{j=1}^{N_{\rms, i}} w_{ij}^{\rm ls}}.
		\label{eq:ESD}
	\end{equation}
	The first sum goes over all lenses and the second sum over all sources separated by the appropriate separation $r_\rmp$ from the lens. The weight $w^\rml$ accounts for the spectroscopic incompleteness of the BOSS galaxy sample,
	\begin{equation}
	    w^\rml = w_{\rm CP} + w_{\rm NOZ} - 1.
	\end{equation}
	Here $w_{\rm CP}$ and $w_{\rm NOZ}$ are weights for the individual lenses that account for fibre collisions and redshift failures, respectively, as detailed in \citet{Anderson_12}. The second weight accounts for the signal-to-noise ratio of each individual lens-source pair and is computed via
	\begin{equation}
		w_{ij}^{\rm ls} = \Sigma_{\mathrm{crit}, ij}^{-2} w_{{\rm lensfit}, i}
	\end{equation}
	with $w_{{\rm lensfit}}$ being the {\sc LENSFIT} weight assigned to each source galaxy that accounts for both the noise in the photometric shape measurement and the ``shape noise'' of intrinsic ellipticies of the galaxy population \citep{Miller_13}. Finally, we follow \cite{Miller_13} and apply an average, ensemble-weighted correction $\langle 1 + m(\nu_{\rm SN}, r) \rangle^{-1}$ to the ESD estimator, where the multiplicative error $m$ depends on the signal-to-noise ratio $\nu_{\rm SN}$ and the galaxy radius $r$ \citep{Miller_13}. This correction factor accounts for inaccuracies in the measured ellipticies and has the strongest deviation from unity for galaxies with low signal-to-noise photometry and small angular sizes.

	We measure the ESD for exactly the same galaxy samples as for the clustering. In all cases, we measure the signal in $11$ logarithmically spaced bins in $r_\rmp$ going from $0.1 \Mpch$ to $15 \Mpch$. We choose $0.1 \Mpch$ as the lower end because we want to avoid scales at which the stellar mass of the lens galaxy dominates the signal \citep{More_15}. Finally, to obtain uncertainties on the galaxy-galaxy lensing signal, we divide the CFHTLenS fields into $30$ roughly equal-area patches and use the jackknife re-sampling technique. We have tested for systematics by measuring the lensing signal around random points and around BOSS galaxies but with the shear rotated by $45\deg$. In both cases we observe a signal consistent with $0$, similar to \cite{Leauthaud_17}, indicating no systematic shear patterns.
	
	The measured galaxy-galaxy lensing signal for all galaxies above $10^{11} \, \Msun$ in stellar mass are shown in the lower panel of Fig.~\ref{fig:hod_fits}. We find a roughly redshift-independent lensing signal, in agreement with \citet{Leauthaud_17} who found this for the smaller redshift range $0.46 < z < 0.7$. The amplitude of the lensing signal is in excellent agreement with previous studies of BOSS galaxies \citep{Miyatake_15, Leauthaud_17}. Finally, we find that the lensing signal for the high stellar mass sample ($11.5 < \log M_\star / \Msun < 12.0$) is higher than for the low stellar mass one by roughly a factor of two, in qualitative agreement with \citet{Miyatake_15}.
	
	\section{\texorpdfstring{$\Lambda$}{Lambda}CDM predictions}
	\label{sec:model}
	
	We start our analysis using an analytical modelling framework. Such an analytical model has the advantage of easily probing the cosmology dependence of the clustering and lensing of galaxies in the BOSS survey. For this analytical model we follow closely the procedure described in \cite{vdBosch_13} and outline the most salient points here. While the accuracy of this approach suffices for the current level of observational constraints, direct mock population \citep[see. e.g.][]{Zheng_16, Hearin_17a, Sinha_18} and emulators \citep[see. e.g.][]{Wibking_19, DeRose_19, Zhai_19} are likely to be preferred for the next generation of surveys.
	
	\subsection{Dark matter haloes}
	\label{sec:DMhaloes}
	
	The model relies on the properties and abundance of dark matter haloes as a function of their mass $M_\rmh$. Here, the halo mass is defined with respect to $200$ times the mean matter density of the Universe. The halo mass function $n_\rmh (M_\rmh)$ and the linear halo bias $b(M_\rmh)$ are computed using the fitting functions of \cite{Tinker_10}. For the linear matter power spectrum, which is needed as an input, we use the transfer function of \cite{Eisenstein_98}. The critical threshold for spherical collapse is assumed to be $0.15 (12\pi)^{2/3} \Omega_\rmm^{0.0055}(z)$ \citep{Navarro_97}. All dark matter haloes are assumed to follow an NFW profile \citep{Navarro_96}. We determine the concentration parameter $c(M_\rmh, z)$ of the NFW profile according to the recipe described in \cite{Bullock_01}, using the updated parameters of \cite{Maccio_07}. We allow for an additional multiplicative factor $1 + \eta$ for all concentration parameters to account for uncertainties in the mass--concentration relation in the literature as well as the scatter in concentration at fixed mass. In particular, \cite{vdBosch_13} showed that setting $\eta = 0.07$ roughly accounts for not directly treating the (log-normal) scatter, which is of the order of $\sigma_{\rm ln c} \simeq 0.3$ \citep{Maccio_07}. In our analysis, we will treat $\eta$ as a nuisance parameter and adopt a Gaussian prior with a mean of 0.07 and a dispersion of 0.05.
	
	\subsection{Halo occupation distribution}
	\label{subsec:hod}
	
	We assume that all galaxies are hosted by dark matter haloes. Furthermore, at least for the analytical model, we postulate that the occupation of haloes with galaxies depends on the halo mass $M_\rmh$ only. Depending on the application, we either use a halo occupation distribution (HOD) or a conditional stellar mass function (CSMF) approach. Here, we first describe the HOD model. As usual, we split the occupation into a central and a satellite galaxy part. Each halo can at most host one central galaxy and an unlimited number of satellites galaxies. The average number of centrals in a halo of mass $M$ is assumed to follow
	\begin{equation}
		\langle N_\rmc | M_\rmh \rangle = \frac{f_\Gamma}{2} \left[ 1 + \mathrm{erf} \left( \frac{\log M_\rmh - \log M_{\rm min}}{\sigma_{\log M_\rmh}} \right) \right].
	\end{equation}
	Here, $f_\Gamma$, $M_{\rm min}$ and $\sigma_{\log M_\rmh}$ are free parameters. With the above parameterisation, $\langle N_\rmc | M_\rmh \rangle$ goes to zero for low halo masses and increases towards higher halo masses. $f_\Gamma$ allows for an overall incompleteness in the target selection of BOSS, i.e. $\langle N_\rmc | M_\rmh \rangle \rightarrow f_\Gamma$ as $M_\rmh \rightarrow \infty$. The average number of satellites is assumed to obey the following parameterisation,
	\begin{equation}
		\langle N_\rms | M_\rmh \rangle = \left( \frac{M_\rmh - M_0}{M_1} \right)^{\alpha}.
	\end{equation}
	$M_0$, $M_1$ and $\alpha$ are free parameters. Additionally, we assume that the satellite numbers follow a Poisson distribution, which implies that
	\begin{equation}
		\langle N_\rms (N_\rms - 1) | M_\rmh \rangle = \langle N_\rms | M_\rmh \rangle^2.
	\end{equation}
	The overall number density of galaxies is obtained via an integral over the halo mass function,
	\begin{equation}
		n_{\rm gal} = \int\limits_0^\infty (\langle N_\rmc | M_\rmh \rangle + \langle N_\rms | M_\rmh \rangle) \ n_\rmh (M_\rmh) \ \rmd  M_\rmh.
	\end{equation} 
	Our HOD model has a total of 6 free parameters, which are listed in Table~\ref{tab:priors} together with their prior ranges that we adopt throughout this study.
	
	\subsection{Conditional stellar mass function}
	
	In cases where we use stellar masses of galaxies as additional observables, we use a CSMF approach \citep{Yang_03, Yang_07, Yang_08, Yang_09, Guo_18, Guo_19}. The CSMF $\Phi (M_\star | M_\rmh)$ describes the average number of galaxies $\rmd N$ with stellar mass in the range $\log M_\star \pm \rmd \log M_\star / 2$ in a halo of mass $M_{\rm h}$. The CSMF of centrals follows a log normal distribution,
	\begin{equation}
		\Phi_\rmc (M_\star | M_\rmh) = \frac{1}{\sqrt{2\pi} \sigma_{\log M_\star}}
		\exp \left\{ - \frac{\log^2 \left[ M_\star / \widetilde{M}_\star (M_\rmh) \right]}{2 \, \sigma_{\log M_\star}^2} \right\},
	\end{equation}
	where
	\begin{equation}
		\widetilde{M}_\star (M_\rmh) = M_{\star, 0} \, \frac{\left( M / M_{\rm h, 1} \right)^{\gamma_1}}{\left( 1 + M / M_{\rmh, 1} \right)^{\gamma_1 - \gamma_2}}
		\label{eq:shmr}
	\end{equation}
	is the median stellar mass of a central galaxy in a halo of mass $M_\rmh$. The variables $\sigma_{\log M_\star}$, $M_{\star, 0}$, $M_{\rmh, 1}$, $\gamma_1$ and $\gamma_2$ are free parameters. We model the CSMF of satellites with a modified \citet{Schechter_76} function,
	\begin{align}
		\Phi_\rms (M_\star | M_\rmh) = &\ln 10 \times \phi_\rms (M_\rmh) \times \left( \frac{M_\star}{0.562 \, \widetilde{M}_\star (M_\rmh)} \right)^{\alpha_s + 1} \times \nonumber\\
		&\exp \left[ - \left( \frac{M_\star}{0.562 \, \widetilde{M}_\star (M_\rmh)} \right)^2 \right],
	\end{align}
	where the normalisation is parameterised according to
	\begin{equation}
		\log \phi_\rms (M_\rmh) = b_0 + b_1 \log M_{12} + b_2 \log M_{12}^2
	\end{equation}
	with $M_{12} = M_\rmh / 10^{12} \Msunh$ and $\alpha_s$, $b_0$, $b_1$ and $b_2$ being free parameters. Finally, we model stellar mass dependent incompleteness effects $\Gamma (M_\star)$ in BOSS according to
	\begin{equation}
		0 \leq \Gamma (M_\star) = \frac{f_\Gamma}{2} \left[ 1 + \mathrm{erf} \left( \frac{\log M_\star - \log M_\Gamma}{\sigma_\Gamma} \right) \right] \leq 1,
	\end{equation}
	with $f_\Gamma$, $M_\Gamma$ and $\sigma_\Gamma$ to be constrained by the data. We follow \citet{Guo_18} and assume separate values of $M_\Gamma$ for centrals and satellites \citep{Guo_18}, which we indicate by $M_{\Gamma, \rmc}$ and $M_{\Gamma, \rms}$, respectively. On the other hand, $f_\Gamma$ and $\sigma_\Gamma$ are the same for centrals and satellites. The average number of centrals and satellites in the stellar mass range $[M_{\star, \rm min}, M_{\star, \rm max}]$ hosted by a dark matter halo of mass $M_\rmh$ can then be obtained via
	\begin{align}
		\langle N | M_\rmh \rangle = \int\limits_{\log M_{\star, \rm min}}^{\log M_{\star, \rm max}} & [ \Gamma_\rmc (M_\star) \, \Phi_\rmc (M_\star | M_\rmh) \, + \, \nonumber \\
		& \, \Gamma_\rms (M_\star) \, \Phi_\rms (M_\star | M_\rmh) ] \, \rmd \log M_\star.
	\end{align}
	and the total number density of galaxies in this stellar mass range simply follows from
	\begin{equation}
      n_{\rm gal}( M_{\star, \rm min}, M_{\star, \rm max}) = \int_0^\infty \langle N | M_\rmh \rangle \, n_\rmh(M_\rmh) \, \rmd M_\rmh\,.
	\end{equation}
    The CSMF model has a total of 13 free parameters, which are listed in Table~\ref{tab:priors} together with their prior ranges that we adopt throughout this study. Note that although the CSMF model has more than twice as many parameters as the HOD model, it describes the halo occupation statistics for {\it any} stellar mass bin, whereas the 6 parameters of the HOD model only characterise one particular mass-threshold sample.

	\subsection{Correlation functions}
	\label{sec:corrfunc}
	
	We express the halo--halo correlation function $\xi_{\rm hh}$ of haloes with masses $M_1$ and $M_2$ via
	\begin{equation}
		\xi_{\rm hh} (r) =
		\left\{\begin{array}{ll}
			\xi_{\rm mm} (r) \, \zeta(r) \, b(M_1) \, b(M_2) & \text{for } r > r_{\rm excl}\\
			-1 & \text{for } r \leq r_{\rm excl}
		\end{array}\right..
		\label{eq:xi_hh}
	\end{equation}
    Here
	\begin{equation}
		r_{\rm excl} = r_{\rm excl} (M_1, M_2) = \mathrm{max}[r_{200m}(M_1), r_{200m}(M_2)]
	\end{equation}
	is the minimum separation between two halo centres and accounts for halo exclusion\citep[see][for details]{vdBosch_13}. Note that $r_{\rm excl}$ is equal to the maximum of the radii of the two halos in question, which are defined as the radii that enclose an average density equal to 200 times the mean background density of the Universe.

	$\xi_{\rm mm}$ is the non-linear matter--matter correlation function obtained via a Fourier transform of the non-linear matter power spectrum, for which we use the fitting formula of \citet{Smith_03}, and $\zeta$ is the modified radial bias function of \citet{vdBosch_13}:
	\begin{equation}
		\zeta(r) =
		\left\{\begin{array}{ll} \zeta_0(r) & \text{for } r \geq r_{\psi}\\
			\zeta(r_\psi)& \text{for } r < r_{\psi}
		\end{array}\,\right.
	\end{equation}
    Here 
	\begin{equation}
		\zeta_0(r) = \frac{\left[1 + 1.17 \xi_{\rm mm} (r)\right]^{1.49}}{\left[1 + 0.69 \xi_{\rm mm} (r)\right]^{2.09}}\,,
	\end{equation}
    is the radial bias function of \citet{Tinker_05}, and $r_{\psi}$ is the root of
	\begin{equation}
		\log \left[ \zeta_0(r_\psi) \, \xi_{\rm mm} (r_\psi)\right] = \psi\,.
	\end{equation}
	The free `nuisance' parameter $\psi$ was introduced by \cite{vdBosch_13} to be able to marginalise over uncertainties in the radial bias function arising from subtle difference in the way dark matter haloes are defined in numerical simulations. As shown in that study, for our definition of halo mass, $\psi \sim 0.9$ which we adopt as our fiducial value in what follows. When marginalising over $\psi$, we instead adopt a Gaussian prior centred on $0.9$ and with a dispersion of 0.15 (see Table~\ref{tab:priors}).
	
	To compute galaxy correlation functions, we split its contribution into a one- and two-halo term,
	\begin{equation}
		\xi_{\rm gx} (r) = \xi_{\rm 1h, gx} (r) + \xi_{\rm 2h, gx} (r),
	\end{equation}
	where $\xi_{\rm gx}$ can stand for the galaxy--galaxy ($\xi_{\rm gg}$) or galaxy--matter ($\xi_{\rm gm}$) correlation function.
	
	The one-halo terms are computed using
	\begin{align}
		\xi_{\rm 1h, gg} (r) = \frac{1}{n_{\rm gal}^2}\sum\limits_{i = \rm c, s} \sum\limits_{\substack{k = \rm c, s\\ik \neq cc}} \int\limits_0^\infty & \left[u_i (M_\rmh) \circledast u_k (M_\rmh) \right] (r) \nonumber\\
		& \langle N_i | M_\rmh \rangle \ \langle N_k | M_\rmh \rangle \ \rmd M_\rmh
	\end{align}
	for the galaxy--galaxy correlation function and
	\begin{align}
		\xi_{\rm 1h, gm} (r) = \frac{1}{n_{\rm gal} \bar{\rho}_\rmm} \sum\limits_{i = \rm c, s}
		\int\limits_0^\infty & \left[ u_i (M_\rmh) \circledast u_\rmm (M_\rmh) \right] (r) \nonumber\\ &\langle N_i | M_\rmh \rangle \ \rmd M_\rmh
	\end{align}
	for the galaxy--matter correlation function. Here, $\circledast$ denotes the three-dimensional convolution operator, $n_{\rm gal}$ is the average number density of galaxies, and $\bar{\rho}_\rmm$ the mean matter density. Both expressions include sums over the central and satellite components. $u_\rmc(r|M)$, $u_\rms(r|M)$ and $u_\rmm(r|M)$ describe the normalised, radial profiles of centrals, satellites and matter, respectively. We assume that centrals are located at the dark matter halo centre, i.e. $u_\rmc(r|M)$ is a three-dimensional Dirac delta function. Satellites, on the other hand, are assumed to follow an NFW profile with a concentration parameter $c_\rms (M_\rmh) = c (M_\rmh) / \mathcal{R}_\rms$ that is different from the matter concentration parameter $c (M_\rmh)$. In our analysis we will treat $\mathcal{R}_\rms$ as a nuisance parameter with a uniform, uninformative prior over the range $[0.5,1.5]$.
	
	In a similar fashion, the two-halo term for the galaxy--galaxy correlation function is
	\begin{align}
		\xi_{\rm 2h, gg} (r) =& \frac{1}{n_{\rm gal}^2} \sum\limits_{i = \rm c, s} \sum\limits_{k = \rm c, s} \int\limits_0^\infty \rmd M_1 \int\limits_0^\infty \rmd M_2  \langle N_i | M_1 \rangle \, \langle N_k | M_2 \rangle \nonumber \\
		& \left[ \xi_{\rm hh}(M_1,M_2) \circledast u_i (M_1) \circledast u_k (M_2) \right] (r) \,,
	\end{align}
	and for the galaxy--matter correlation function is
	\begin{align}
		\xi_{\rm 2h, gm} (r) =& \frac{1}{n_{\rm gal} \bar{\rho}_\rmm} \sum\limits_{i = \rm c, s} \int\limits_0^\infty \rmd M_1 \int\limits_0^\infty \rmd M_2 \langle N_i | M_1 \rangle \, M_2 \nonumber \\
		& \left[ \xi_{\rm hh}(M_1,M_2) \circledast u_i (M_1) \circledast u_\rmm (M_2) \right] (r) \,,
	\end{align}
	
	$w_\rmp$ and $\Delta\Sigma$ are computed via a LOS integration of the galaxy--galaxy and galaxy--matter correlation functions, respectively. In doing so, we also correct $w_\rmp$ for residual redshift-space distortions\footnote{They arise from the fact that the projected correlation function for the data is computed by integrating only out to a maximum LOS distance of $r_{\pi,{\rm max}} = 100 \Mpch$ (see \S\ref{sec:clustering}).} using the recipe described in \cite{vdBosch_13}. We note that the accuracy of the entire analytic model has been tested against numerical simulations in \cite{vdBosch_13}. Nevertheless, we present an additional test in appendix \ref{sec:analytic_accuracy}.
	
	Finally, whenever we adopt a cosmology other than the Planck18 cosmology defined at the end of Section~\ref{sec:intro}, we correct the predicted observables for the fact that the measurements presented in Section~\ref{sec:data} have made use of the distance-redshift relation for the Planck18 cosmology. In particular, we correct $n_{\rm gal}$ and $w_\rmp$ using the method of \cite{More_13a}. The correction of $\Delta\Sigma$, however, is non-trivial and depends on the source galaxy redshift distribution. Generally, the corrections are only at the level of a few percent, which is smaller than the typical observational uncertainties, and we therefore neglect this complication for $\Delta\Sigma$.
	
	\section{Lensing discrepancy}
	\label{sec:discrepancy}
	
    Here, we present the observational results for $w_\rmp$ and $\Delta \Sigma$ obtained in Section \ref{sec:data} and analyse the relation between the predicted and observed lensing signal. Specifically, we will show that models for the galaxy-halo connection tuned to reproduce the clustering overpredict the lensing signal. In addition, we will investigate the redshift and stellar mass dependence of this lensing discrepancy.
	
	\subsection{Dependence on redshift}
	
	\begin{table}
		\centering
		\begin{tabular}{ccc}
			\hline
			Type & Parameter & Prior\\
			\hline\hline
			HOD & $\log M_{\rm min}$ & $[10.0, 15.0]$\\
			HOD & $\log M_0$ & $[10.0, 15.0]$\\
			HOD & $\log M_1$ & $[10.0, 16.0]$\\
			HOD & $\sigma_{\log M_\rmh}$ & $[0.2, 1.0]$\\
			HOD & $\alpha$ & $[0.1, 3.0]$\\
			HOD & $f_\Gamma$ & $[0.5, 1.0]$\\
			\hline\hline
			CSMF & $\log M_{\star, 0}$ & $[9.0, 11.5]$\\
			CSMF & $\log M_{\rmh, 1}$ & $[10.0, 14.0]$\\
			CSMF & $\gamma_1$ & $[2.0, 5.0]$\\
			CSMF & $\gamma_2$ & $[0.1, 0.7]$\\
			CSMF & $\sigma_{\log M_\star}$ & $[0.1, 0.2]$\\
			CSMF & $b_0$ & $[-2.5, 0.5]$\\
			CSMF & $b_1$ & $[0.0, 2.0]$\\
			CSMF & $b_2$ & $[-0.5, 0.5]$\\
			CSMF & $\alpha_{\rm s}$ & $[-3.0, 0.0]$\\
			CSMF & $f_\Gamma$ & $[0.0, 1.0]$\\
			CSMF & $\sigma_\Gamma$ & $[0.0, 1.0]$\\
			CSMF & $\log M_{\Gamma, \rmc}$ & $[10.0, 12.0]$\\
			CSMF & $\log M_{\Gamma, \rms}$ & $[10.0, 12.0]$\\
			\hline\hline
			Nuisance &$\psi$ & $\mathcal{N}(0.9, 0.15)$\\
			Nuisance &$\eta$ & $\mathcal{N}(0.07, 0.05)$\\
			Nuisance &$\mathcal{R}_\rms$ & $[0.5, 1.5]$\\
			\hline
		\end{tabular}
		\caption{The priors used in the analysis in Section \ref{sec:discrepancy}. $[a, b]$ indicates flat priors and $\mathcal{N}(\mu, \sigma)$ a normal distribution with mean $\mu$ and scatter $\sigma$.}
		\label{tab:priors}
	\end{table}
	
	We first investigate the redshift dependence of the ESD and compare it to the predicted values based on the clustering. To this end, we fit an HOD model to the abundance and $w_\rmp$ for all galaxies with stellar masses above $10^{11} \Msun$ in the $6$ redshift bins. The prior ranges are listed in Table \ref{tab:priors}. The HOD model parameters in each redshift bin are allowed to be completely independent. We fix the cosmological parameters to the best-fit values of the Planck18 analysis. The nuisance parameters in the model are set to $\psi = 0.9$, $\eta = 0.07$ and $\mathcal{R}_\rms = 1$. The motivation for the values of $\psi$ and $\eta$ are given in Sections~\ref{sec:corrfunc} and~\ref{sec:DMhaloes}, respectively, while setting $\mathcal{R}_\rms = 1$ implies that we assume that satellite galaxies are an unbiased tracer of the mass distribution of their dark matter host halo.
	
	We use {\sc MultiNest} to fit the $6$ free HOD parameters in each redshift bin. Here, and throughout this paper, when running {\sc MultiNest} we employ $1000$ live points, a target efficiency of $0.1$ and $\Delta \ln \mathcal{Z} = 0.01$ as the stopping criterion. Here, $\mathcal{Z}$ is the estimate for the global evidence \citep{Feroz_13}. For completeness, we list all posteriors on the galaxy-halo connection in Table \ref{tab:gal_posteriors}.
	
    \begin{figure*}
		\centering
		\includegraphics[width=\textwidth]{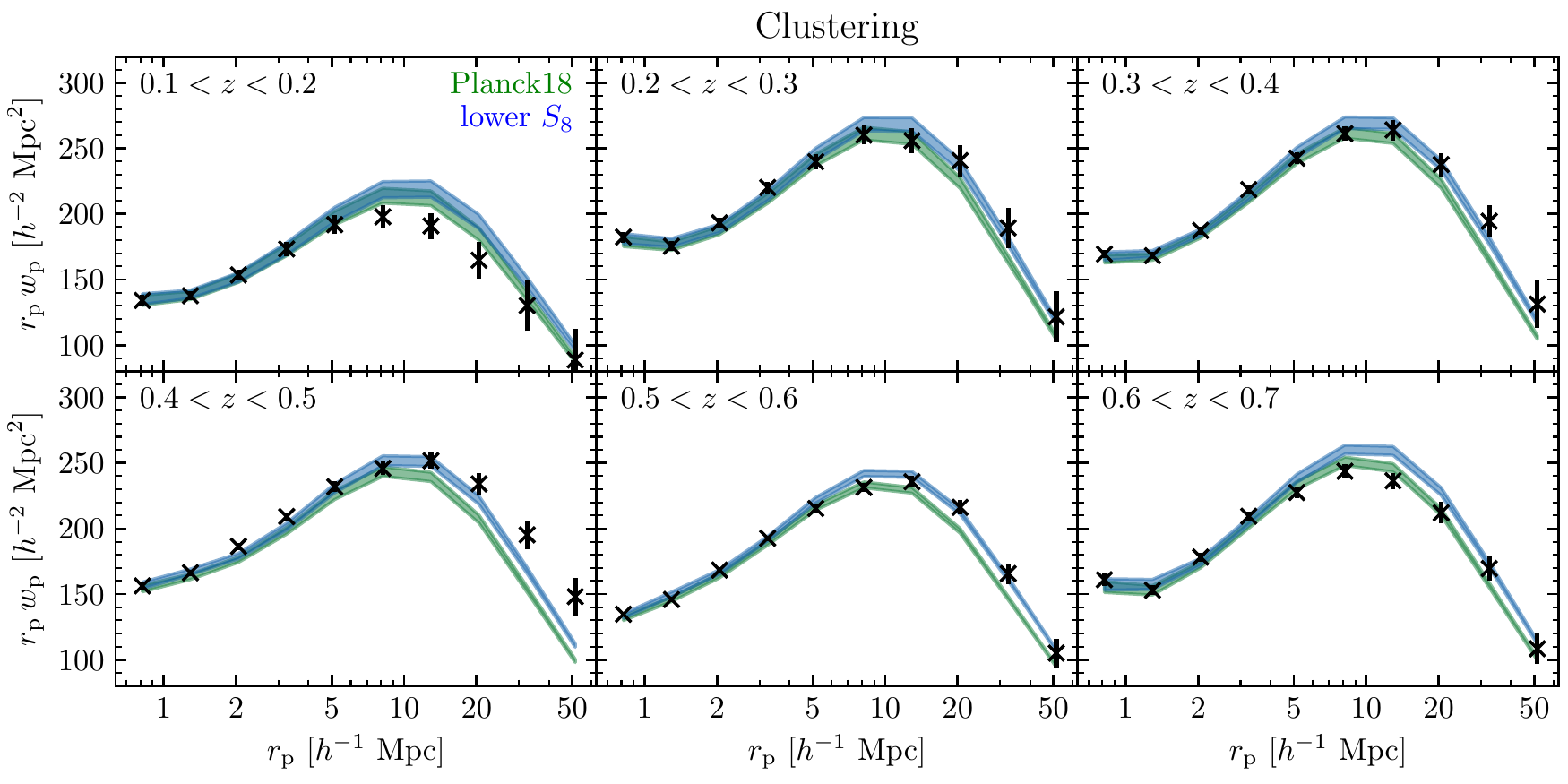}
		\includegraphics[width=\textwidth]{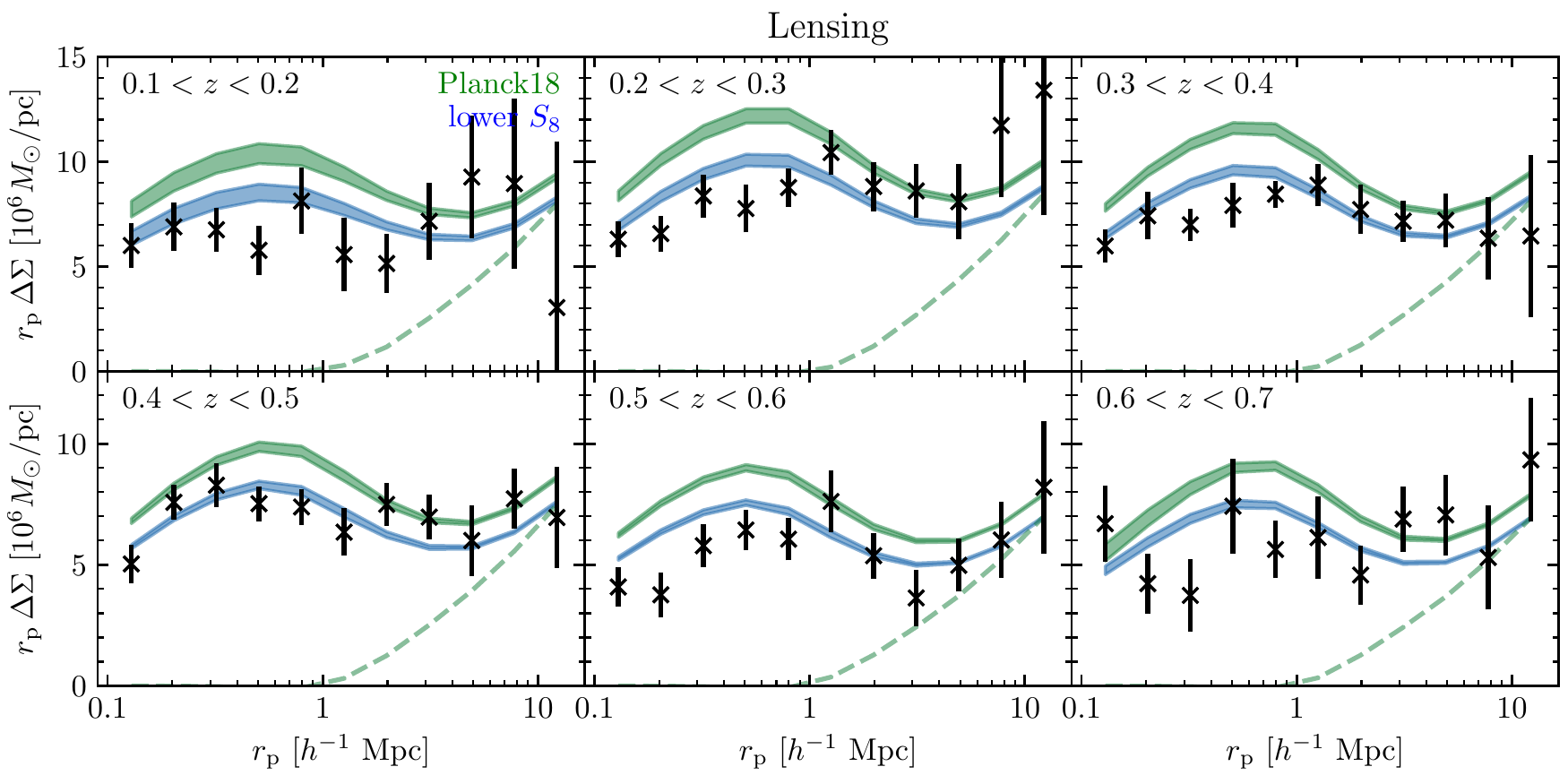}
		\caption{The clustering (top panel) and lensing signal (bottom panel) of BOSS galaxies for different redshifts. In all cases, we consider BOSS galaxies with stellar masses above $10^{11} \Msun$. Error bars denote measurements with $1\sigma$ errors. Bands show $68\%$ posterior predictions of HOD fits assuming either the Planck18 cosmological parameters (green) or Planck18 parameters but with $\Omega_\rmm = 0.3$ and $\sigma_8 = 0.75$ (blue). Note that the models have been fit to the abundance and $w_\rmp$ only. On the other hand, $\Delta\Sigma$ is a pure prediction. Finally, for $\Delta\Sigma$, we also show the median prediction coming entirely from the 2-halo term as the dashed line.}
		\label{fig:hod_fits}
	\end{figure*}
	
	The results are presented in  Fig.~\ref{fig:hod_fits}, which shows the measurements for $w_\rmp$ (upper panels) and $\Delta\Sigma$ (lower panels) as error bars, while the green shaded bands show the $68\%$ confidence intervals from our posterior probability distributions. Note that the excess surface densities were not used as constraints on the models; in the case of $\Delta\Sigma$ the green bands are therefore purely (posterior) predictions from the models fit to $n_{\rm gal}$ and $w_\rmp$ only. By eye, our models fit the observed $w_\rmp$ reasonably well. However, the $\chi^2$ per degree of freedom (dof) values are generally unlikely with $\chi^2 / \mathrm{dof} \sim 3 - 8$. Some of this might be attributed to inaccuracies in the analytical model, specifically the corrections for residual redshift space distortions. Generally, we do not expect the modelling inaccuracies to significantly impact our predictions for $\Delta\Sigma$ because the relative uncertainties in $w_\rmp$ are an order of magnitude smaller than those of $\Delta\Sigma$. Additionally, the analytical model does not allow for the possibility of galaxy assembly bias. As discussed below, this might also contribute to the poor fit.
	
	We now investigate the predicted and observed lensing signal. As shown in the lower panel of Fig.~\ref{fig:hod_fits}, we find that on small scales, $r_\rmp < 3 \Mpch$, the predicted lensing signal is systematically higher than the observed one. The discrepancy is generally of the order of $\sim 30\%$. Furthermore, the overprediction of the lensing signal for Planck18 cosmological parameters is present and consistent for all redshifts in the range $0.1 < z < 0.7$. We have verified that marginalising over the nuisance parameters using the priors listed in Table \ref{tab:priors} does not change the conclusions. The discrepancy is slightly mitigated by allowing for lower halo concentrations, i.e. $\eta < 0$. However, even for an extreme value such as $\eta = -0.3$, implying that halo concentrations are 30 percent smaller than for the fiducial model, the problem is only mitigated on scales $r_\rmp < 0.5 \Mpch$ and has little impact on scales $0.5 \Mpch < r_\rmp < 3 \Mpch$.
    
    Having verified that the lensing discrepancy is largely independent of redshift, we now average the result over all redshift bins in order to increase the signal-to-noise ratio. We neglect uncertainties in the predicted lensing signal since they are significantly smaller than the observational uncertainties $\sigma(\Delta\Sigma_{\rm obs})$. Both the mean and uncertainty in each $r_\rmp$ bin comes from a simple error-weighted mean, i.e.
    \begin{equation}
        \left\langle \frac{\Delta\Sigma_{\rm obs}}{\Delta\Sigma_{\rm pred}} \right\rangle = \frac{\sum\limits_{i = 1}^6 \frac{\Delta\Sigma_{{\rm obs}, i}}{\Delta\Sigma_{{\rm pred}, i}} \left( \frac{\sigma (\Delta\Sigma_{{\rm obs}, i})}{\Delta\Sigma_{{\rm pred}, i}} \right)^{-2}}{\sum\limits_{i = 1}^6 \left( \frac{\sigma (\Delta\Sigma_{{\rm obs}, i})}{\Delta\Sigma_{{\rm pred}, i}} \right)^{-2}}
    \end{equation}
    and
    \begin{equation}
        \sigma \left\langle \frac{\Delta\Sigma_{\rm obs}}{\Delta\Sigma_{\rm pred}} \right\rangle = \sqrt{\sum\limits_{i = 1}^6 \left( \frac{\sigma (\Delta\Sigma_{{\rm obs}, i})}{\Delta\Sigma_{{\rm pred}, i}} \right)^{-2}}^{-1} \, .
    \end{equation}
    In Fig.~\ref{fig:esd_ratio}, we show in green the average ratio of the predicted to observed lensing signal. The left-hand panel shows the ratio for the full sample analysed thus far (i.e., with $\log M_\star / \Msun > 11$). From this plot it is apparent that the discrepancy between observed and predicted lensing signal is limited to smaller scales, $r_\rmp < 3 \Mpch$, whereas the signal is consistent with the prediction at larger scales, albeit with large uncertainties.
	
	\subsection{Dependence on stellar mass}
	
	In the previous subsection, we analysed the lensing signal for effectively the entire BOSS galaxy sample as a function of redshift. We now divide the sample into a low ($11.0 \leq \log M_\star < 11.5$) and high stellar mass ($11.5 \leq \log M_\star < 12.0$) one and measure the clustering and lensing amplitudes for both samples. The motivation is that more massive galaxies live in more massive dark matter haloes on average. Thus, we expect different clustering and lensing signals from both samples. Compared to the previous section, we now fit the CSMF model to the data in each redshift bin. Again, we use the analytical model and only fit the abundance via the (incomplete) stellar mass function and the two $w_\rmp$ measurements for both samples, not $\Delta\Sigma$. The adopted minima and maxima for the flat CSMF priors are listed in Table \ref{tab:priors} and the nuisance parameters are kept fixed.
	
	\begin{figure*}
		\centering
		\includegraphics[width=\textwidth]{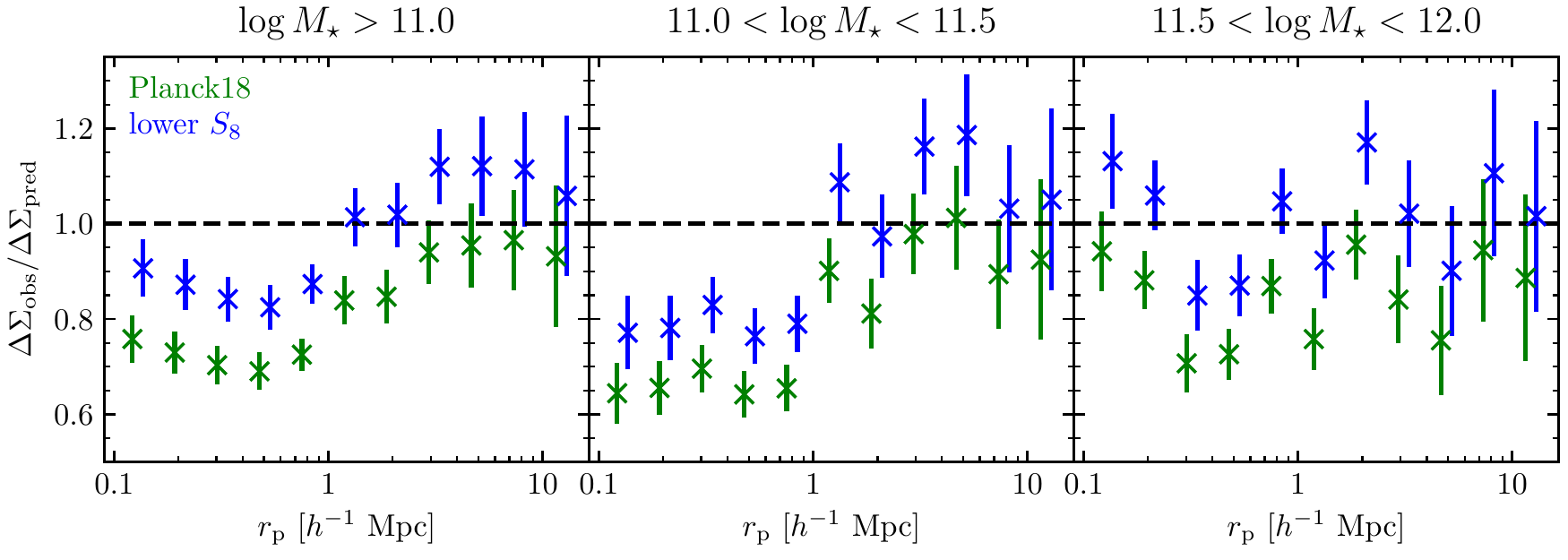}
		\caption{Ratio of the predicted to observed lensing signal averaged over all redshifts, $0.1 < z < 0.7$. Different panels correspond to samples of different stellar masses as indicated in the title. Predictions are made using the Planck18 cosmological parameters (green) and for a cosmological model with lower $S_8$ than Planck18 but otherwise identical parameters (blue).}
		\label{fig:esd_ratio}
	\end{figure*}
	
	\cite{Guo_18} carried out a very similar analysis by fitting a CSMF model to the number density and clustering of BOSS galaxies in different redshift bins, including the same split in stellar mass. However, there are a few key differences compared to our approach here. First, the authors use results from the BigMDPL simulation \citep{Klypin_16} instead of an analytic model to predict $n_{\rm gal}$ and $w_\rmp$. Secondly, \cite{Guo_18} assume that satellites follow the same galaxy stellar--halo mass relation (SHMR) as centrals, i.e. Eq.~\eqref{eq:shmr}. In this case, the halo mass of satellites is taken to represent the mass at accretion onto the host halo, i.e. $M_{\rm acc}$. Altogether, this implies that their model has less freedom for the satellite occupation. Also, the authors do not fit the scatter in stellar mass at fixed halo mass, but rather assume $\sigma_{\log M_\star} = 0.173$. Reassuringly, we are able to reproduce the most salient trends of their analysis. In particular, we infer very similar completeness levels: $f_\Gamma \sim 0.9$ at $0.1 < z < 0.5$ and $f_\Gamma \sim 0.5$ at $0.5 < z < 0.6$. Similarly, our results suggest that $\log M_{\Gamma, \rmc}$ is generally larger than $\log M_{\Gamma, \rms}$, indicating that satellites at a given stellar mass are more likely to pass the BOSS selection cut than centrals. This is likely due to the fact the satellites, on average, are redder than centrals \citep[see e.g.][]{Xu_18}. Also, with the exception of the $0.4 < z < 0.5$ bin, we infer very similar values for $\gamma_2$, the slope of the SHMR at the high-mass end. However, we find the slope at the low mass end, $\gamma_1$, to be virtually unconstrained within our adopted prior range $2.0 < \gamma_1 < 5.0$, whereas \cite{Guo_18} find $\gamma_2 \sim 8$. Generally, our findings for the low-mass slope are in much better agreement with previous literature results \citep{Yang_09, More_11, Leauthaud_12, Behroozi_13b}. Finally, we infer a stellar mass scatter of $\sigma_{\log M_\star} \sim 0.15 - 0.2$, again in good agreement with other studies \citep{Yang_09, More_11, Leauthaud_12, Saito_16} and the value assumed in \cite{Guo_18}.
	
	Overall, we find that our CSMF inferences based on clustering are in good agreement with other independent studies. We now turn our attention again to the galaxy-galaxy lensing signal. As expected, we find that both the measured and predicted lensing amplitudes are higher for the high stellar mass sample in all redshift bins. This reflects that more massive galaxies live in more massive dark matter haloes. Also, in agreement with the results from the previous subsection, we find no significant redshift evolution for either the predicted or observed lensing signal for both samples. In the middle and right panel of Fig.~\ref{fig:esd_ratio}, we show the redshift-averaged ratio of the predicted to the observed lensing signal for the two samples. The low stellar mass sample reveals the same discrepancy on small scales ($r_\rmp \lta 3 \Mpch$) as for the full sample, albeit somewhat noisier due to the reduced sample size. For the high mass sample, the results are even noisier, but there is still a clear indication that the model overpredicts the lensing signal, with $\Delta\Sigma_{\rm obs} / \Delta\Sigma_{\rm pred} = 0.85 \pm 0.02$ for $r_\rmp < 3 \Mpch$. On the other hand, on larger scales, the predictions are consistent with the measurements for both samples. Hence, we conclude that the lensing discrepancy found for the entire BOSS sample is also present in the two subsamples split by stellar mass.

	\section{Scrutinizing the discrepancy}
	\label{sec:explaining}

    Thus far, we have established that our model for the galaxy-dark matter connection is unable to simultaneously fit the clustering plus galaxy-galaxy lensing data for BOSS galaxies, if we adopt a standard (`vanilla') $\Lambda$CDM cosmology with Planck18 parameters. The model, when fit to the clustering data, consistently and systematically overpredicts the lensing data on small scales. We now examine three potential solutions to this discrepancy: changing cosmology, galaxy assembly bias, and the impact of baryonic feedback on the matter distribution.

    \subsection{Cosmology}
    \label{subsec:cosmology}

    The combination of galaxy clustering and galaxy-galaxy lensing is a sensitive probe of cosmology \citep{Seljak_05, Yoo_06, Cacciato_09, More_13b, Leauthaud_17}, especially $\Omega_\rmm$ and $\sigma_8$. In order to gauge to what extent the discrepancy highlighted in the previous section depends on these cosmological parameters, we repeat the same analysis after changing $(\Omega_\rmm, \sigma_8) = (0.3153, 0.8111)$, which are the Planck18 values, to $(0.3,0.75)$. This reduces the value of $S_8 \equiv \sigma_8 \sqrt{\Omega_\rmm / 0.3}$ from $0.8315$ to $0.75$. All other cosmological parameters are left unchanged. The resulting model {\it fits} to $w_\rmp$, as well as the corresponding model {\it predictions} for $\Delta\Sigma$, are shown in blue in Figures \ref{fig:hod_fits} and \ref{fig:esd_ratio}. Especially from Fig.~\ref{fig:esd_ratio} it is apparent that the reduced value for $S_8$ results in a significant decrease in the predicted galaxy-galaxy lensing signal, bringing it in much better agreement with the data. Note, though, that the difference with respect to predictions for the Planck18 cosmology are roughly independent of scale and redshift. Consequently, while such a reduced value for $S_8$ can now satisfactorily explain $\Delta\Sigma$ on small scales, the lensing signal is under-predicted on large scales. Hence, it appears that a simple change in cosmological parameters is unable to fully explain the discrepancy.

    \begin{table}
        \centering
        \begin{tabular}{lcc}
            \hline
            Parameter & Prior & Posterior\\
            \hline\hline
            $\Omega_\rmm$ & $[0.20, 0.35]$ & $0.2966_{-0.0050}^{+0.0053}$\\
            $\sigma_8$ & $[0.5, 1.0]$ & $0.748_{-0.017}^{+0.017}$\\
            $\Omega_\rmb h^2$ & $\mathcal{N}(0.02237, 0.00015)$ & $0.02246_{-0.00014}^{+0.00013}$\\
            $n_s$ & $\mathcal{N}(0.9649, 0.0042)$ & $0.9681_{-0.0039}^{+0.0039}$\\
            $h$ & $\mathcal{N}(0.6736, 0.0054)$ & $0.6783_{-0.0053}^{+0.0051}$\\
            \hline\hline
            $\psi$ & $\mathcal{N}(0.9, 0.15)$ & $0.931_{-0.096}^{+0.131}$\\
            $\eta$ & $\mathcal{N}(0.07, 0.05)$ & $-0.066_{-0.040}^{+0.040}$\\
            $\mathcal{R}_\rms$ & $[0.5, 1.5]$ & $1.36_{-0.19}^{+0.10}$\\
            \hline
		\end{tabular}
		\caption{The priors and posteriors for the cosmological and nuisance parameters when analysing all measurements at all redshift simultaneously. The priors for $\Omega_\rmb h^2$, $n_s$ and $h$ come from the Planck18 analysis. We also take into account the covariance between those three parameters (not shown).}
		\label{tab:cosmo}
	\end{table}

	To further investigate this issue, we now fit all parameters (including the cosmological and nuisance parameters listed in Table~\ref{tab:priors}) simultaneously to the  $n_{\rm gal}$, $w_\rmp$ and $\Delta\Sigma$ measurements at all redshifts.	In each of the $6$ redshift bins, we analyse the measurements for the full $\log(M_\star / M_\odot) > 11$ sample, i.e., no split in stellar mass is applied. For each set of cosmological and nuisance parameters, we find the $6$ HOD models that maximise the likelihood in each of the $6$ redshift bins. The total log-likelihood for this combination of cosmological and nuisance parameters is then set to the sum of the best-fit log-likelihoods in each redshift bin. As shown in Table \ref{tab:cosmo}, we use Planck18 priors on $H_0$, $\Omega_\rmb h^2$ and $n_s$, including the covariance among them, and flat priors for $\Omega_\rmm$ and $\sigma_8$. The HOD and nuisance parameters assume the same priors as in the previous section.
	
	\begin{figure}
		\centering
		\includegraphics[width=\columnwidth]{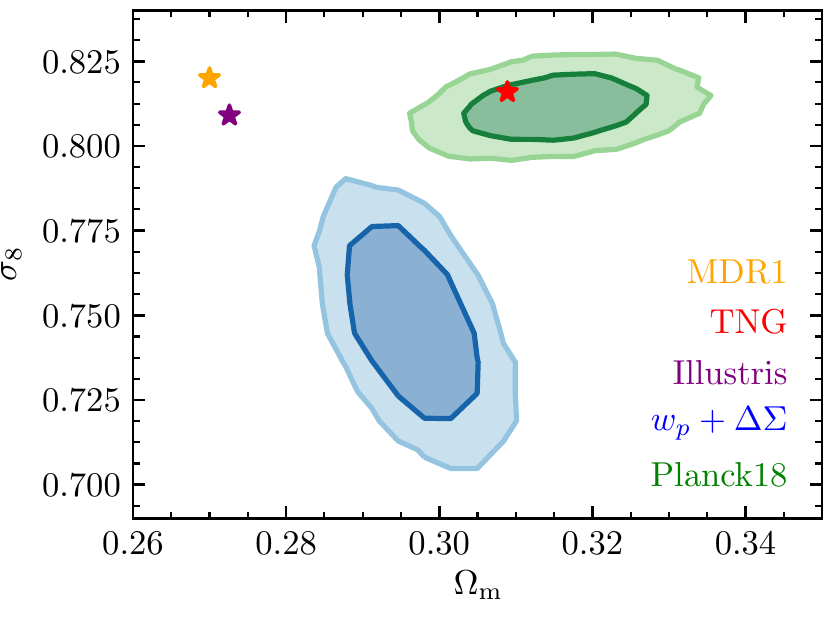}
		\caption{Constraints on $\Omega_\rmm$ and $\sigma_8$ when fitting $n_{\rm gal}$, $w_\rmp$ and $\Delta\Sigma$ in $6$ redshift bins spanning $0.1 < z < 0.7$ (blue). Flat priors for $\Omega_\rmm$ and $\sigma_8$ are assumed, whereas the priors on $H_0$, $\Omega_\rmb h^2$ and $n_s$ come from the Planck18 CMB analysis. We also show the posteriors for $\Omega_\rmm$ and $\sigma_8$ from Planck18 (green). Bands denote the $68\%$ and $95\%$ posterior ranges.}
		\label{fig:Om0_sigma8}
	\end{figure}
	
	The one-dimensional posteriors and best-fit values for the cosmological and nuisance parameters are listed in Table \ref{tab:cosmo}. The posterior constraints on $H_0$, $\Omega_\rmb h^2$ and $n_s$ are driven by the Planck18 priors. Thus, in Fig.~\ref{fig:Om0_sigma8} we concentrate on the constraints on $\Omega_\rmm$ and $\sigma_8$. We also show the constraints from the Planck18 analysis. If the lensing discrepancy described in the previous sections could be ascribed entirely to a change in cosmology, our results would imply a significantly lower value for $S_8 = \sigma_8 \sqrt{\Omega_\rmm / 0.3}$ than what is found by Planck18. Specifically, our posterior is centred on $\Omega_\rmm = 0.3$ and $\sigma_8 = 0.75$, the parameter combination we tested previously. Overall, we infer $S_8 = 0.744_{-0.015}^{+0.015}$, whereas Planck18 finds $0.832_{-0.013}^{+0.013}$, a discrepancy of $4.4\sigma$. However, given the unsatisfactory fit presented in Fig. \ref{fig:esd_ratio} it appears that a change of cosmological parameters, within the 6-parameter `vanilla' $\Lambda$CDM cosmology cannot fully explain the small-scale lensing discrepancy. We leave the exploration of additional freedom in cosmological parameters, such as a non-zero neutrino mass or more freedom in the Hubble parameter, for future investigations.
	
	\subsection{Galaxy assembly bias}
	\label{subsec:galaxy_assembly_bias}
	
	An assumption inherent in the modelling framework in section \ref{sec:model} is that dark matter halo mass is the only variable governing the occupation of haloes with galaxies. If dark matter halo mass were the only variable determining the clustering of haloes, this simplification would not impact the predictions for $\Delta\Sigma$ at fixed $w_\rmp$. However, it is well known that the clustering of dark matter haloes depends on other variables besides mass \citep{Gao_05, Wechsler_06, Villarreal_17}, an effect called halo assembly bias. Thus, if galaxy occupation correlates with any other of these halo variable besides mass, an effect called galaxy assembly bias, predictions for $w_\rmp$ and $\Delta\Sigma$ are impacted. Ultimately, neglecting assembly bias in the modelling can lead to inferences that are systematically and significantly biased \citep{Zentner_14}. This effect should be particularly strong for small-scale $\Delta\Sigma$ measurements that probe the dark matter halo structure, but can also have a non-negligible impact on intermediate scales \citep[][]{Sunayama_16}.
	
	\subsubsection{Simulation}
	
    Unfortunately, there is no analytic framework for predicting the clustering properties of galaxies and its cosmological dependence in the presence of galaxy assembly bias. Thus, we cannot use the modelling framework presented in section \ref{sec:model} and instead turn to results obtained from cosmological simulations. Essentially, we will repeat the exercise of fitting a model for the galaxy-halo connection to the clustering $w_\rmp$ and predict the lensing signal $\Delta\Sigma$, this time allowing for galaxy assembly bias. The disadvantage of using a simulation for this is that cosmological parameters cannot be changed. In this work, we use results from the MultiDark MDR1 simulation \citep{Prada_12}. MDR1 is a dark matter-only N-body simulations that traces $(2048)^3$ particles in a cosmological volume with $1 \Gpch$ on a side. We use the publicly available halo catalogue at $z = 0.53$ derived with the ROCKSTAR \citep{Behroozi_13a} halo finder. The cosmological parameters for MDR1 are $\Omega_\rmm = 0.27$, $\sigma_8 = 0.82$, $h = 0.7$, $n_s = 0.95$ and $\Omega_\rmb h^2 = 0.023$. We populate dark matter haloes in the MultiDark simulation using the {\sc halotools} (v0.6) package \citep{Hearin_17a}. Centrals are assumed to be at rest with respect to the phase-space position of the halo and satellites are assumed to follow an NFW profile. We assume a minimum halo mass of $300$ times the particle mass, $300 m_\rmp = 2.6 \times 10^{12} \Msunh$, to host a galaxy. This is more than adequate for modelling the occupation statistics of our population of BOSS galaxies, which all have a stellar mass $M_\star \geq 10^{11} \Msunh$.
    
    We also use {\sc halotools} to generate mock catalogues and predict the different observables. The correlation functions $w_\rmp$ and $\Delta\Sigma$ are computed by using the distant observer approximation and projecting the galaxy and matter distribution along each of the three axes of the simulation. We randomly down-sample the full particle distribution in MDR1 by a factor of $200$ in order to reduce computational cost when computing $\Delta \Sigma$. We do not use the default {\sc delta\_sigma} function of {\sc halotools} v0.6 to compute $\Delta\Sigma$ and instead implement our own routine. This new algorithm is described and motivated in appendix \ref{sec:new_delta_sigma}.
	
    \subsubsection{Decorated HOD}
	
	In addition to the standard HOD parameterisation described in section \ref{subsec:hod}, we also test the \textit{decorated} HOD (dHOD) framework \citep{Hearin_16}. In this model, the occupation is allowed to depend on a secondary halo parameter at fixed mass. Thus, it is a model that allows for galaxy assembly bias. We split the haloes into two populations based upon whether the secondary halo parameter is above or below the median at that halo mass. The occupation with centrals is then given by
	\begin{equation}
	    \langle N_\rmc | M_\rmh, x \rangle = \langle N_\rmc | M_\rmh \rangle \pm A_{\rm cen} \left( 0.5 - \left| 0.5 - \langle N_\rmc | M_\rmh \rangle \right| \right),
	\end{equation}
	where $\pm$ depends on whether the value of the secondary halo parameter $x$ is above or below the median. $\langle N_\rmc | M_\rmh \rangle$ is the standard, mass-only occupation in the HOD model described previously and $A_{\rm cen}$ is a free parameter in the range $[-1, +1]$. Similarly, the occupation with satellites is given by
	\begin{equation}
	    \langle N_\rms | M_\rmh, x \rangle = (1 \pm A_{\rm sat})\langle N_\rms | M_\rmh \rangle
	\end{equation}
	with $A_{\rm sat}$ being another free parameter spanning the range $[-1, +1]$. The model reduces to our default HOD model without assembly bias when $A_{\rm cen} = A_{\rm sat} = 0$, while assembly bias is maximised for $\left| A_{\rm cen} \right|$ and $\left| A_{\rm sat} \right|$ being unity. The decorated HOD is defined such that a non-zero $A_{\rm cen}$ or $A_{\rm sat}$ preserves the mean occupation of the default HOD parameters; the galaxies are merely redistributed among haloes of the same mass. Note, though, that since we continue to assume that $N_{\rm sat}$ obeys Poisson statistics, with a mean given by $\langle N_\rms | M_\rmh, x \rangle$, the expectation value for the number of satellite-satellite pairs for a given halo mass, $\langle N_\rms (N_\rms - 1) | M_\rmh \rangle$, is not preserved when $A_{\rm sat} \ne 0$  \citep[see][for a detailed discussion]{Hearin_16}.
	
    In order to efficiently predict $w_\rmp$ and $\Delta \Sigma$ for a given dHOD model, we have implemented and generalized the framework presented in \cite{Zheng_16} of tabulating correlation functions as a function of halo properties, i.e. halo mass $M_\rmh$ and a secondary halo property $x$. This bypasses the need to populate dark matter haloes directly and dramatically speeds up the computation of $w_\rmp$ and $\Delta \Sigma$ as a function of dHOD parameters. We have tested that it gives the same results as the direct mock population method used, for example, in \cite{Zentner_19}.
    
    \subsubsection{Results}
    
    \begin{figure}
		\centering
		\includegraphics[width=\columnwidth]{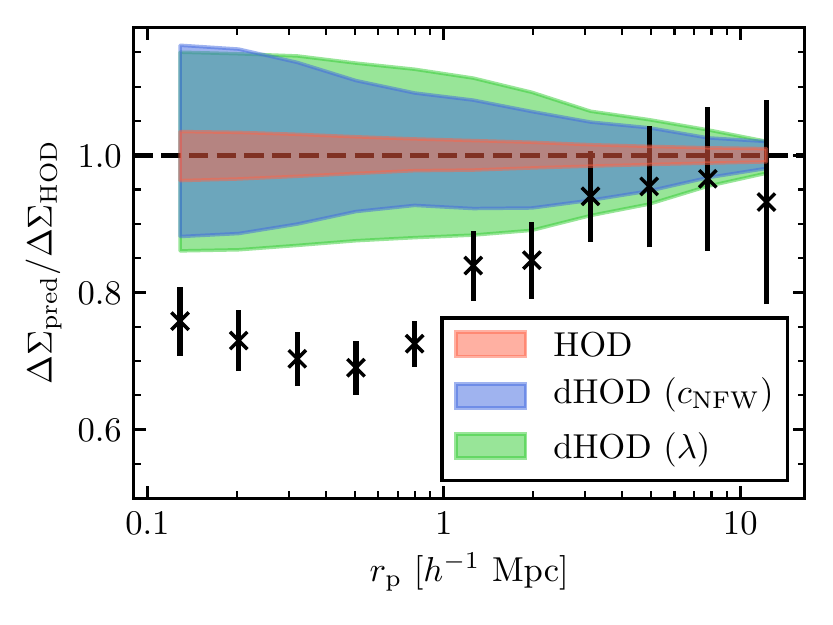}
		\caption{Ratio of the posterior predictions for the galaxy-galaxy lensing signal from fitting $w_\rmp$ for the $0.5 < z < 0.6$ sample. We divide all predictions by the median value for the HOD prediction. Bands always denote the $95\%$ posterior range. Both $w_\rmp$ and $\Delta\Sigma$ have been predicted from the MultiDark simulation. We show results for three different parameterisations: the traditional HOD and two dHOD models with concentration $c_{\rm NFW}$ and spin $\lambda$ as secondary halo parameters. To guide the eye, we also include the ratio of observed to predicted lensing signal from the left-hand panel of Fig.~\ref{fig:esd_ratio}.}
		\label{fig:esd_hod_vs_dhod}
	\end{figure}
	
	In the following, we will estimate the potential of galaxy assembly bias to explain the small-scale lensing discrepancy. We repeat the exercise of fitting $n_{\rm gal}$ and $w_\rmp$ to observations and predicting $\Delta\Sigma$ but using MDR1 instead of the analytical model. We do this for observations in the redshift range $0.5 < z < 0.6$. We test $3$ different models: the conventional HOD and two dHOD models with halo concentration $c_{\rm NFW}$ and halo spin $\lambda$ as the secondary halo parameter. In Fig.~\ref{fig:esd_hod_vs_dhod} we show the ratio of the different predictions to the median of the standard (no galaxy assembly bias) HOD result. Bands denote the $95\%$ posterior prediction in all cases. We only select models with $f_\Gamma > 0.8$ because the completeness has been found observationally to be higher than $0.8$ \citep{Leauthaud_16}. Taking into account all models with $f_\Gamma > 0.5$ widens the confidence intervals by $\sim 30\%$.
	
	We see that $\Delta\Sigma$ at large scales, $r_\rmp > 10 \Mpch$, is basically unaffected by the choice of the model. This conforms to our naive expectation since $w_\rmp$ fixes the large-scale bias of the galaxy sample and $\Delta\Sigma$ at large $r_\rmp$ mainly depends on this bias, not halo mass. For smaller $r_\rmp$ that probe the one-halo term the uncertainties are larger for the dHOD models. This is expected since assembly bias affects the relationship between halo mass and bias, allowing for different average halo masses at fixed clustering strength. We find that the small-scale $\Delta\Sigma$ predictions are positively and negatively correlated with the assembly bias parameter $A_{\rm cen}$ for $c_{\rm NFW}$ and $\lambda$ as secondary halo parameters, respectively. In other words, we expect lower values for $\Delta\Sigma$ on small scales if BOSS galaxies preferentially occupy low-concentration or high-spin haloes. Such a scenario would alleviate the lensing discrepancy alluded to above.
	
	Overall, we find strong evidence that galaxy assembly bias could alleviate the lensing discrepancy. However, even in the $95\%$ posterior range, we only find a decrease of at most $10\%$ for the small-scale lensing signal whereas the observed difference is of the order of $25\%$. Furthermore, we find that $w_\rmp$ alone places no strong constraints on the exact values of the assembly bias parameters \citep[also see][]{Vakili_19, Zentner_19}. In fact, even values of $|A_{\rm cen}| = 1$, i.e. maximum assembly bias strength, cannot be strongly excluded and contribute to the posterior in Fig.~\ref{fig:esd_hod_vs_dhod}. Thus, at face value, it seems difficult to explain the entire discrepancy as arising from galaxy assembly bias. However, we note that we have only tested one rather specific model for galaxy assembly bias as presented in \cite{Hearin_16}. Ultimately, a more comprehensive study of the impact of galaxy assembly bias, based on more generic models, is required before we can rule out assembly bias as the root cause of the lensing discrepancy.
	
    \subsection{Baryonic feedback}
    \label{subsec:baryonic_feedback}
    
    So far, our results were implicitly or explicitly based on placing galaxies into dark matter haloes in collisionless N-body simulations. However, those simulations neglect the effect of baryons on the overall matter distribution in the Universe. Specifically, it has been shown that baryonic effects can alter the matter distribution on large enough scales relevant to our work \citep[see e.g.][]{Jing_06, Rudd_08}. Thus, we turn to the results of hydrodynamical simulations in order to test the impact of baryonic feedback. Unfortunately, in addition to not being able to vary cosmology, those simulations are generally too expensive to be run on volumes comparable to the BOSS galaxy survey. Thus, we cannot marginalise over uncertainties of the galaxy-halo connection by comparing predictions to observations. However, we will show below that the relative impact of baryonic physics on $\Delta\Sigma$ is mostly insensitive of the properties of the tracer galaxy population. We thus still expect our results to hold generally.
    
    \subsubsection{Simulations}
    
    We use results from two simulation suites: Illustris \citep{Vogelsberger_14, Nelson_15} and TNG300 \citep{Pillepich_18, Springel_18, Nelson_18, Naiman_18, Marinacci_18, Nelson_19}. For both  simulation suites, we use the highest resolution versions. The Illustris-1 simulation traces structure and galaxy formation in a cubic volume of $\left( 75 \Mpch \right)^3$ and achieves a mass resolution of $6.3 \times 10^6 \ \Msun$ and $1.3 \times 10^6 \ \Msun$ for dark matter and baryons, respectively. The TNG300-1 simulation has a volume of $\left( 205 \Mpch \right)^3$ and a mass resolution of $5.9 \times 10^7 \ \Msun$ and $1.1 \times 10^7 \ \Msun$. Both Illustris-1 and TNG300-1 also have dark matter only versions with the same initial conditions but no baryonic physics. We will use these in comparison with their hydrodynamic couterparts to test the impact of baryonic physics. From both simulation suites, we use the $z = 0.55$ snapshots and the subhalo catalogues derived with {\sc subfind}. Finally, the cosmological parameters of Illustris are derived from the WMAP9 analysis \citep{Hinshaw_13}, i.e. $S_8 = 0.77$, while TNG300 uses cosmological parameters compatible with the Planck15 analysis \citep{Planck_16}, i.e. $S_8 = 0.83$.

    To compute $\Delta\Sigma$, we again use {\sc{halotools}} and a down-sampled version of the full particle distribution. However, the baryonic runs of the simulations already contain massive black hole particles which we want to avoid to downsample further. Thus, instead of a random down-sampling, we use the following algorithm: For each particle with mass $m_\rmp$ we calculate $f = m_\rmp / m_\rmt$, where $m_\rmt$ is some target mass. If $f \geq 1$, i.e. $m_\rmp \geq m_\rmt$, the particle will automatically be in the down-sampled catalogue. If $f < 1$, the particle will be in the down-sampled catalogue with a probability of $f$ and assigned a mass of $m_\rmp / f = m_\rmt$. When computing $\Delta \Sigma$, we use target masses of $m_\rmt = 10^{9} \ \Msunh$ and $m_\rmt = 10^{10} \ \Msunh$ for Illustris and TNG300, respectively.
    
    \subsubsection{Matching haloes}
    
    In each of the two simulation suites, we first cross-match dark matter field haloes in the full-physics run to the dark matter-only run. We ignore subhaloes because such a matching is much more challenging and because satellite galaxies have a negligible contribution to the overall lensing signal. For Illustris-1, a halo matching between the dark matter and the full physics run already exists in the online database. For TNG300-1 we match field haloes by requiring a field halo in the dark matter-only run within $0.15 \Mpch \, (M_{\rm vir} / 10^{13} \Msunh)^{1/3}$, where $M_{\rm vir}$ is the virial mass of the halo in the baryonic run. This matching radius corresponds to roughly half the virial radius. In a few cases this leads to clearly spurious matches with low-mass haloes in the dark matter-only run. We therefore require that the mass difference between the haloes in the matched pairs is less than $1.0 \ \mathrm{dex}$. Overall, for both simulations, we can match upwards of $\sim 95\%$ of all haloes.
    
    \subsubsection{Results}
    
    \begin{figure}
		\centering
		\includegraphics[width=\columnwidth]{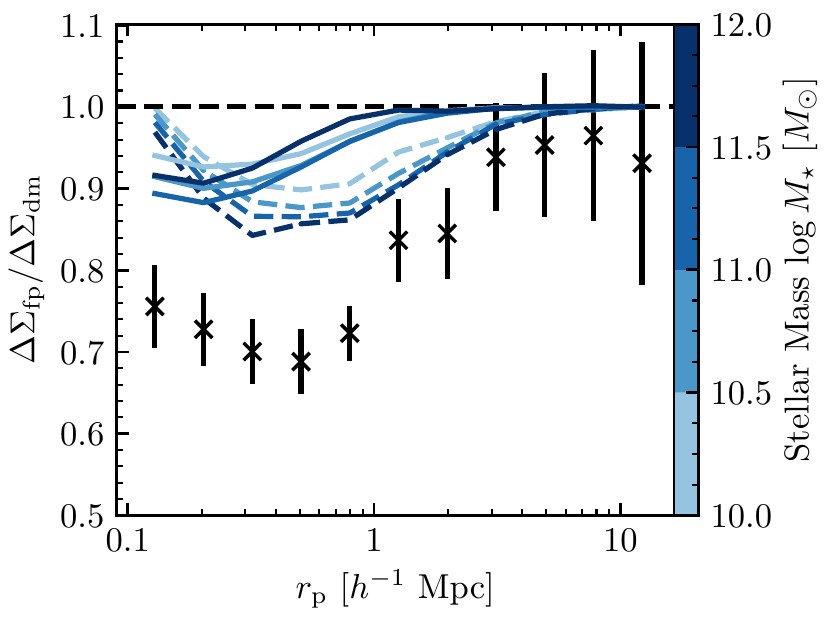}
		\caption{The impact of baryons on the expected lensing signal as probed by TNG300 (solid lines) or Illustris (dashed lines). We show the ratio of the ESD in the full physics run to the dark matter-only run for matched field haloes. The colours indicate the stellar mass of the hosted galaxy in the full physics run, as indicated by the colour bar on the right. The effect of baryonic physics is significantly less pronounced in TNG300 as compared to Illustris. We again include the ratio of observed to predicted lensing signal from the left-hand panel of Fig.~\ref{fig:esd_ratio}.}
		\label{fig:esd_baryon_ratio}
	\end{figure}
    
    In the next step, we select all field haloes in the full physics run of the simulation that host galaxies of a given stellar mass and have a match in the dark matter-only run. We use the {\sc delta\_sigma} routine of {\sc halotools} to compute the expected galaxy-galaxy lensing signal. In all cases, we use the down-sampled particle catalogue and project the particle and galaxy distribution onto the three spatial axes separately. The total lensing signal is then taken to be the average of the three projections. We show the ratio of the lensing signal in the full physics run to the signal around matched haloes in the dark matter-only run in Fig.~\ref{fig:esd_baryon_ratio}. In general, both simulations predict a change in the galaxy-galaxy lensing signal due to baryonic physics. On large scales, ($r_\rmp > 3 \Mpch$), both simulations predict a negligible impact. Instead, the effects of baryonic feedback start to become more important on smaller scales.
    
    For Illustris, this happens already at scales of around $3 \Mpch$, whereas for TNG this happens at $1 \Mpch$. Generally, the impact of baryonic physics is significantly larger in Illustris, reaching up to $15\%$, than in TNG, where it reaches up to $10\%$. We attribute this difference primarily to the different treatments of AGN feedback \citep{Weinberger_17, Springel_18}. The implementation in TNG300 produces galaxy and intercluster medium (ICM) properties that are in much better agreement with observational constraints \citep{Weinberger_18}. Interestingly, we also find that the scale-dependence and relative importance of baryonic feedback for galaxy-galaxy lensing is only a weak function of the stellar mass of the host galaxy. For Illustris, the relative importance decreases slightly with stellar mass, whereas in TNG300 it has a maximum for stellar masses around $10^{11} \ \Msun$.

    Overall, the impact of baryonic physics goes in the right direction of decreasing the lensing signal on small scales, while not affecting larger scales. However, given that this decrease is at most $10\%$ in the case of TNG300, it seems challenging to resolve the entire lensing discrepancy by baryonic effects. Nevertheless, it is apparent that constraining baryonic feedback effects is crucial in order to derive unbiased cosmological constraints from small-scale lensing.
	
	\section{Comparison with previous cosmological studies}
	\label{sec:cosmology_comparison}

    The findings of this study are particularly relevant for the possible tension regarding $S_8$ between studies of the low-redshift Universe and the CMB constraints. Generally, our results are in broad agreement with previous cosmological studies modelling the clustering and lensing of galaxies down to small scales.
    
    \cite{Leauthaud_17} find that $\Delta\Sigma$ is overpredicted when using pure dark-matter only simulations based on the Planck15 cosmological parameters. Our findings extend those of \cite{Leauthaud_17} which are based on the entire CMASS sample at $z \sim 0.5$ by showing that this result is insensitive to changes in redshift and stellar mass. We also show that models fit to the BOSS clustering as a function of redshift predict a roughly redshift-independent lensing signal. As discussed in \cite{Leauthaud_17}, this stands in contrast to the naive expectations based on subhalo abundance matching by \cite{Saito_16} and might indicate redshift-dependent biases in the CMASS stellar mass estimates or complicated sample selection functions. \cite{Leauthaud_17} also discuss in detail systematic effects besides a change in cosmological parameters that could lead to the low lensing signal, most importantly baryonic effects. They analyse data from the Illustris simulation and obtain results that are qualitatively in agreement with what we present in section \ref{subsec:baryonic_feedback}, i.e. the lensing signal is reduced by baryonic feedback by around $15\%$ on small scales, $0.1 \Mpch < r_\rmp < 1 \Mpch$. Similarly, under the assumption that the lensing discrepancy can be ascribed entirely to a change in cosmology, their constraints on $\Omega_\rmm$ and $\sigma_8$ are compatible with ours, but with much larger uncertainties.

    \begin{figure}
        \centering
        \includegraphics[width=\columnwidth]{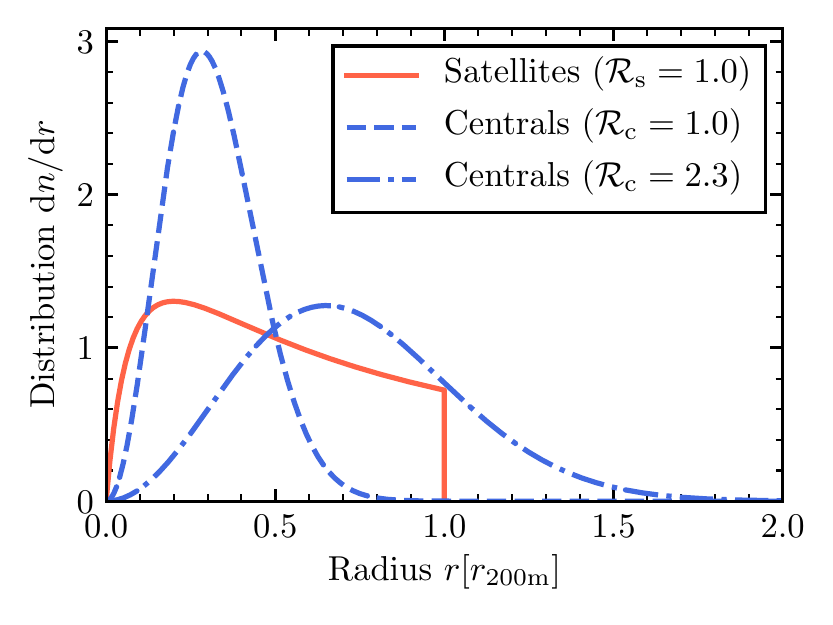}
        \caption{The radial distribution of galaxies in a halo with concentration parameter $c_{200} = 5$. The distributions are normalized to unity and the mass or redshift of the halo are irrelevant. We show the distribution of satellites following an NFW profile with $R_\rms = 1.0$ (red, solid), centrals with $\mathcal{R}_\rmc = 1.0$ (blue, dashed) and centrals with $\mathcal{R}_\rmc = 2.3$ (blue, dash-dotted).}
		\label{fig:radial_distribution}
	\end{figure}
    
    \cite{More_15} also derive cosmological parameters through an analysis of the clustering and lensing of the BOSS CMASS sample. They find constraints on $\Omega_\rmm$ and $\sigma_8$ that are roughly in between and in agreement with both the results of Section \ref{subsec:cosmology} and those of Planck18. We speculate that part of the reason for their agreement with Planck18 is that \cite{More_15} allow for the off-centring of central galaxies. Specifically, \cite{More_15} assume that a certain fraction of centrals are not located at the halo centre but instead follow a Gaussian distribution,
    \begin{equation}
        u_\rmc (r | M_\rmh) = \frac{1}{(2 \pi)^{3/2} \mathcal{R}_\rmc^3 r_\rms^3 (M_\rmh)} \exp \left( - \frac{r^2}{2 \mathcal{R}_\rmc^2 r_\rms^2 (M_\rmh)} \right)\,.
    \end{equation}
    \cite{More_15} infer that 30 to 40 per cent of all centrals follow such a distribution with  $\mathcal{R}_\rmc \simeq 2.3$ (i.e., with a scale-radius more than double that of the dark matter halo). We find that such an off-centring is extremely efficient in decreasing the predicted lensing signal at the smallest scales, $r_\rmp < 1 \Mpch$. Specifically, for those centrals that are off-centred with such a large value of $\mathcal{R}_\rmc$, the ESD, $\Delta\Sigma$, decreases by $\sim 50\%$ ($\sim 100 \%$) at $0.8 \Mpch$ ($0.1 \Mpch$). Thus, assuming that $30\%$ of all centrals have such an offset is equivalent to reducing the galaxy-galaxy lensing signal on small scales by roughly the same amount. Hence, in principle such an off-centring could potentially explain the observed lensing discrepancy. However, as we show in  Fig.~\ref{fig:radial_distribution}, this implies that roughly one third of all centrals have halo-centric positions that are less centrally concentrated than satellite galaxies. This clearly is an extreme scenario, for which there is no substantial support. In fact, such large off-centring should be accompanied by a large intra-halo velocity dispersion of centrals \citep{vdBosch_05a, Skibba_11, Lange_18}, at odds with the inferred low central velocity dispersion in BOSS CMASS \citep{Reid_14, Guo_15a}. In addition, hydrodynamical simulations by \cite{Cui_16} typically show offsets of less than $\sim 5\%$ the scale radius between the brightest cluster galaxy (BCG) and the halo centre. Less than $10\%$ of all clusters in the simulations show BCG offsets that exceed $0.1 \, r_{200 \rmm}$, and a large fraction of those are likely to be cases in which the BCG is actually a satellite galaxy \citep[see e.g.][]{Skibba_11, Hoshino_15, Lange_18}. Hence, in the simulations of \cite{Cui_16}, at most a few per cent of all ``centrals'' have offsets comparable to $\mathcal{R}_\rmc \simeq 1.0$. To summarise, we find no observational or theoretical argument that would support the large off-centring assumed in \cite{More_15}.
	
	Cosmological constraints using small-scale clustering and galaxy-galaxy lensing have also been obtained from the SDSS main galaxy sample by \cite{Cacciato_13}. Their constraints on $\Omega_\rmm$ and $\sigma_8$ are in excellent agreement with our results in Fig.~\ref{fig:Om0_sigma8}, and thus in tension with Planck18. By extension, this suggest that a lensing discrepancy similar to what we observe in BOSS is also present in the lower redshift SDSS data \citep[also see][]{Zu_15, Zu_16, Behroozi_18}. We emphasize, though, that \cite{Cacciato_13}, similar to \cite{More_15}, did not account for either assembly bias or the potential impact of baryonic physics.
	
	As shown in section \ref{sec:explaining}, the effects of baryonic feedback and galaxy assembly bias vanish at larger scales. The recent \textit{Dark Energy Survey} (DES) analysis of galaxy clustering, galaxy-galaxy lensing and cosmic shear concentrates on such large scales and also finds lower values for $\Omega_\rmm$ and $\sigma_8$ compared to Planck18, albeit somewhat less acute \citep{DES_18a}. The same is true for the clustering and lensing analyses of \cite{Mandelbaum_13} using SDSS galaxies and \cite{Singh_18} using the LOWZ sample.  Finally, several recent cosmic shear analyses, including \textit{Hyper Suprime-Cam} \citep[HSC, ][]{Hikage_19}, the \textit{Deep Lens Survey} \citep[DLS, ][]{Jee_16}, the \textit{Kilo Degree Survey} \citep[KiDS, ][]{Hildebrandt_17} and CFHTLenS \citep{Fu_14}, also find similar results. Thus, there is significant evidence that the true value for $S_8$ is slightly lower than the Planck18 value of $0.83$. But also see \cite{Chang_19} for a unified analysis of those $4$ shear studies and their reliability. Similarly, \cite{Liu_16} provide tentative evidence based on CMB lensing that CFHTLenS shear estimates are systematically too low.

    \section{Conclusion}
    \label{sec:conclusion}

    We have presented a new analysis examining the clustering and galaxy-galaxy lensing signal of galaxies in the BOSS survey. Our analysis extends previous findings by \cite{Miyatake_15}, \cite{More_15} and \cite{Leauthaud_17}. Our main findings are as follows.
	
	\begin{itemize}
	    \item When adopting the Planck18 $\Lambda$CDM cosmology and a model for the galaxy-halo connection fit to the clustering of BOSS galaxies, the ESD is consistently overy-predicted on small scales ($r_\rmp < 3 \Mpch$) by roughly $30 \%$ compared to observations.
	    \item This over-prediction of the lensing signal is independent of the redshift ($0.1 < z < 0.7$, cf. Fig.~\ref{fig:hod_fits}) and the stellar mass ($11 < \log M_\star / M_\odot < 12$, cf. Fig.~\ref{fig:esd_ratio}) of the sample in question.
	    \item A change in cosmology, particularly allowing for a lower values of $S_8 = \sigma_8 \sqrt{\Omega_\rmm / 0.3} \sim 0.75 \pm 0.02$ (cf. Fig.~\ref{fig:Om0_sigma8}) can alleviate the discrepancy. However, the cosmological parameters needed to entirely eliminate the discrepancy are in tension with Planck18 findings. In addition, there are some indications that such a low value for $S_8$ predicts an excess surface density that is too large on {\it large} scales ($r_\rmp > 3 \Mpch$).
	    \item The clustering data alone does not strongly constrain the presence of galaxy assembly bias. Allowing for this effect alleviates, but does not remove, the small-scale lensing discrepancy in BOSS (cf. Fig.~\ref{fig:esd_hod_vs_dhod}).
	    \item We investigated the impact of baryonic physics on the galaxy-galaxy lensing signal in the newly-released IllustrisTNG simulation. Compared to Illustris, we find that baryonic physics lead to a smaller reduction of the lensing signal of at most $10\%$ (cf. Fig.~\ref{fig:esd_baryon_ratio}).
	\end{itemize}
	
	Our study highlights the tension between large-scale structure probes and the CMB regarding cosmological parameters that has also been identified in several previous studies (see Section \ref{sec:cosmology_comparison} for references). However, on the small-scales probed here, a better understanding of galaxy formation physics, in particular galaxy assembly bias and baryonic feedback, is needed to draw firm conclusions regarding the $\Lambda$CDM model. In addition, the recent study by \citet{Chang_19} suggests that there may also be an issue with the reliability of the galaxy-galaxy lensing data, something that should hopefully improve with more and better data coming available. In particular, new and independent galaxy-galaxy lensing measurements from ongoing and future surveys such as DES, HSC and/or LSST should improve both the quantity and quality of the data, allowing for a more rigorous test of our cosmological framework. It will be crucial, though, to complement this forthcoming data with significant improvements in the modelling of clustering and lensing
    \citep[see e.g.][]{Wibking_19, DeRose_19, Zhai_19, Nishimichi_18}. In addition, further advances will come from combining galaxy clustering and galaxy-galaxy lensing with additional, alternative probes of large-scale structure, such as redshift space distortions \citep[e.g.,][]{Yang_08, Reid_14}, satellite kinematics \citep[e.g.,][]{More_11, Lange_19b}, higher-order correlation functions \citep[][]{Gil-Marin_17, Gualdi_19}, cosmic shear \citep[e.g,][]{Fu_14, Hildebrandt_17}, and counts-in-cells \citep[e.g.,][]{Reid_09, Gruen_18}.  Such additional data will prove especially important for breaking degeneracies and constraining galaxy assembly bias \citep[see e.g.,][]{Wang_19}. Ultimately, the tension described here clearly deserves further investigation which will lead to either new insights into galaxy-formation physics or a revision of our cosmological paradigm.
 
	\section*{Acknowledgements}

	We thank Martin White, Alexie Leauthaud and Simone Ferraro for interesting discussions regarding this work.
	
	FvdB and JUL are supported by the US National Science Foundation (NSF) through grant AST 1516962. This research was supported  by the HPC facilities operated by, and the staff of, the Yale Center for Research Computing. FvdB received additional support from the Klaus Tschira foundation, and from the National Aeronautics and Space Administration through Grant No. 17-ATP17-0028 issued as part of the Astrophysics Theory Program. XY, HG and WL are supported by the National Key Basic Research Program of China (Nos. 2015CB857002, 2015CB857003), national science foundation of China (Nos. 11833005, 11890692, 11621303, 11655002, 11773049). HG acknowledges the support of the 100 Talents Program of the Chinese Academy of Sciences. This work is also supported by a grant from Science and Technology Commission of Shanghai Municipality (Grants No. 16DZ2260200).
	
	This work made use of the following software packages: {\sc matplotlib} \citep{Hunter_07}, {\sc SciPy} \citep{Jones_01}, {\sc NumPy} \citep{vdWalt_11}, {\sc Astropy} \citep{Astropy_13}, {\sc Cython} \citep{Behnel_11}, {\sc Corner} \citep{Foreman-Mackey_16}, {\sc MultiNest} \citep{Feroz_08,Feroz_09, Feroz_13}, {\sc PyMultiNest} \citep{Buchner_14}.
	
	The CosmoSim database used in this paper is a service by the Leibniz-Institute for Astrophysics Potsdam (AIP). The MultiDark database was developed in cooperation with the Spanish MultiDark Consolider Project CSD2009-00064.
	
	\bibliographystyle{mnras}
	\bibliography{bibliography}
	
	\appendix
	
	\section{Galaxy-halo model posteriors}
	
	For completeness, we list in Table~\ref{tab:gal_posteriors} all posterior predictions for the galaxy-halo parameters in section \ref{sec:discrepancy} and \ref{subsec:cosmology}. Note that the CSMF parameters in the bin $0.3 < z < 0.4$ show a strong bimodality. In this case, there exists another mode with $\gamma_2 \sim 0.3$ and $M_{\Gamma, \rmc} > M_{\Gamma, \rms}$ that also includes the maximum-likelihood fit.
	
	\begin{table*}
		\centering
		\begin{tabular}{ccccccc}
			\hline
			Parameter & \multicolumn{6}{c}{Posterior}\\
			& $0.1 < z < 0.2$ & $0.2 < z < 0.3$ & $0.3 < z < 0.4$ & $0.4 < z < 0.5$ & $0.5 < z < 0.6$ & $0.6 < z < 0.7$\\
			\hline\hline
			\multicolumn{7}{l}{Type: HOD, Cosmology: Planck18}\\
			\hline
			$\log M_{\rm min}$ & $13.096_{-0.095}^{+0.107}$ & $13.260_{-0.073}^{+0.079}$ & $13.259_{-0.071}^{+0.083}$ & $13.23_{-0.14}^{+0.13}$ & $13.18_{-0.14}^{+0.11}$ & $13.77_{-0.16}^{+0.15}$\\
			$\log M_0$ & $11.67_{-1.16}^{+1.23}$ & $11.55_{-1.05}^{+1.14}$ & $11.67_{-1.14}^{+1.16}$ & $11.41_{-0.96}^{+1.02}$ & $11.46_{-0.99}^{+1.13}$ & $11.65_{-1.13}^{+1.26}$\\
			$\log M_1$ & $14.244_{-0.051}^{+0.039}$ & $14.348_{-0.028}^{+0.026}$ & $14.389_{-0.032}^{+0.027}$ & $14.349_{-0.025}^{+0.024}$ & $14.349_{-0.025}^{+0.023}$ & $14.72_{-0.10}^{+0.24}$\\
			$\sigma_{\log M_{\rm h}}$ & $0.39_{-0.13}^{+0.13}$ & $0.351_{-0.102}^{+0.100}$ & $0.349_{-0.097}^{+0.097}$ & $0.46_{-0.15}^{+0.10}$ & $0.451_{-0.151}^{+0.096}$ & $0.757_{-0.110}^{+0.090}$\\
			$\alpha$ & $1.18_{-0.13}^{+0.12}$ & $1.304_{-0.094}^{+0.082}$ & $1.29_{-0.13}^{+0.10}$ & $1.641_{-0.097}^{+0.093}$ & $1.66_{-0.12}^{+0.10}$ & $1.15_{-0.32}^{+0.29}$\\
			$f_\Gamma$ & $0.83_{-0.12}^{+0.11}$ & $0.850_{-0.098}^{+0.095}$ & $0.843_{-0.097}^{+0.104}$ & $0.77_{-0.13}^{+0.15}$ & $0.79_{-0.13}^{+0.14}$ & $0.75_{-0.17}^{+0.18}$\\
			\hline\hline
			\multicolumn{7}{l}{Type: HOD, Cosmology: lower $S_8$}\\
			\hline
			$\log M_{\rm min}$ & $13.05_{-0.13}^{+0.14}$ & $13.191_{-0.094}^{+0.103}$ & $13.189_{-0.095}^{+0.105}$ & $13.17_{-0.16}^{+0.12}$ & $13.11_{-0.15}^{+0.12}$ & $13.71_{-0.17}^{+0.14}$\\
			$\log M_0$ & $11.57_{-1.06}^{+1.25}$ & $11.49_{-1.03}^{+1.12}$ & $11.50_{-1.01}^{+1.17}$ & $11.30_{-0.90}^{+1.01}$ & $11.40_{-0.95}^{+1.05}$ & $11.63_{-1.12}^{+1.22}$\\
			$\log M_1$ & $14.153_{-0.052}^{+0.039}$ & $14.252_{-0.030}^{+0.026}$ & $14.284_{-0.031}^{+0.025}$ & $14.270_{-0.026}^{+0.025}$ & $14.268_{-0.026}^{+0.025}$ & $14.489_{-0.036}^{+0.068}$\\
			$\sigma_{\log M_{\rm h}}$ & $0.45_{-0.16}^{+0.15}$ & $0.39_{-0.12}^{+0.11}$ & $0.39_{-0.12}^{+0.10}$ & $0.48_{-0.16}^{+0.10}$ & $0.458_{-0.154}^{+0.097}$ & $0.796_{-0.090}^{+0.068}$\\
			$\alpha$ & $1.24_{-0.13}^{+0.11}$ & $1.380_{-0.087}^{+0.084}$ & $1.40_{-0.12}^{+0.10}$ & $1.798_{-0.100}^{+0.098}$ & $1.88_{-0.12}^{+0.12}$ & $1.53_{-0.30}^{+0.25}$\\
			$f_\Gamma$ & $0.78_{-0.14}^{+0.14}$ & $0.80_{-0.12}^{+0.13}$ & $0.80_{-0.11}^{+0.13}$ & $0.78_{-0.16}^{+0.15}$ & $0.78_{-0.14}^{+0.15}$ & $0.74_{-0.17}^{+0.18}$\\
			\hline\hline
			\multicolumn{7}{l}{Type: CSMF, Cosmology: Planck18}\\
			\hline
			$\log M_{\star, 0}$ & $10.911_{-0.087}^{+0.090}$ & $10.782_{-0.105}^{+0.096}$ & $11.266_{-0.163}^{+0.070}$ & $11.249_{-0.102}^{+0.085}$ & $10.99_{-0.12}^{+0.12}$ & $10.63_{-0.30}^{+0.24}$\\
			$\log M_{{\rm h}, 1}$ & $11.85_{-0.19}^{+0.20}$ & $11.72_{-0.20}^{+0.21}$ & $12.35_{-0.18}^{+0.20}$ & $12.37_{-0.15}^{+0.19}$ & $12.11_{-0.17}^{+0.20}$ & $11.65_{-0.42}^{+0.30}$\\
			$\gamma_1$ & $3.58_{-1.00}^{+0.94}$ & $3.53_{-1.03}^{+1.01}$ & $3.61_{-1.01}^{+0.97}$ & $3.61_{-0.98}^{+0.93}$ & $3.57_{-1.02}^{+0.95}$ & $3.48_{-0.95}^{+0.98}$\\
			$\gamma_2$ & $0.217_{-0.028}^{+0.023}$ & $0.314_{-0.022}^{+0.019}$ & $0.149_{-0.034}^{+0.079}$ & $0.165_{-0.039}^{+0.045}$ & $0.311_{-0.041}^{+0.034}$ & $0.438_{-0.045}^{+0.042}$\\
			$\sigma_{\log M_\star}$ & $0.1638_{-0.0091}^{+0.0084}$ & $0.1624_{-0.0095}^{+0.0086}$ & $0.2023_{-0.0092}^{+0.0076}$ & $0.2272_{-0.0079}^{+0.0076}$ & $0.1975_{-0.0113}^{+0.0098}$ & $0.191_{-0.038}^{+0.027}$\\
			$b_0$ & $-0.99_{-0.82}^{+0.61}$ & $-1.59_{-0.59}^{+0.51}$ & $-0.51_{-0.97}^{+0.69}$ & $-0.92_{-0.82}^{+0.74}$ & $-1.27_{-0.76}^{+0.93}$ & $-1.12_{-0.79}^{+0.87}$\\
			$b_1$ & $0.74_{-0.52}^{+0.64}$ & $0.56_{-0.38}^{+0.46}$ & $0.99_{-0.56}^{+0.65}$ & $0.64_{-0.44}^{+0.65}$ & $0.88_{-0.55}^{+0.61}$ & $1.00_{-0.61}^{+0.56}$\\
			$b_2$ & $0.01_{-0.15}^{+0.12}$ & $0.113_{-0.089}^{+0.075}$ & $-0.08_{-0.16}^{+0.18}$ & $0.117_{-0.138}^{+0.098}$ & $-0.03_{-0.21}^{+0.15}$ & $-0.12_{-0.20}^{+0.18}$\\
			$\alpha_{\rm s}$ & $-1.02_{-0.93}^{+0.66}$ & $-0.80_{-0.28}^{+0.31}$ & $-1.44_{-0.99}^{+1.08}$ & $-1.05_{-1.01}^{+0.72}$ & $-0.57_{-1.13}^{+0.44}$ & $-0.96_{-1.03}^{+0.61}$\\
			$f_\Gamma$ & $0.936_{-0.077}^{+0.046}$ & $0.960_{-0.054}^{+0.029}$ & $0.964_{-0.049}^{+0.026}$ & $0.930_{-0.070}^{+0.048}$ & $0.894_{-0.087}^{+0.072}$ & $0.57_{-0.17}^{+0.24}$\\
			$\sigma_\Gamma$ & $0.116_{-0.017}^{+0.019}$ & $0.0842_{-0.0072}^{+0.0072}$ & $0.1539_{-0.0122}^{+0.0071}$ & $0.222_{-0.015}^{+0.014}$ & $0.178_{-0.025}^{+0.014}$ & $0.269_{-0.031}^{+0.020}$\\
			$\log M_{\Gamma, \rm c}$ & $11.116_{-0.025}^{+0.016}$ & $11.2585_{-0.0077}^{+0.0083}$ & $11.241_{-0.011}^{+0.014}$ & $11.209_{-0.019}^{+0.017}$ & $11.223_{-0.019}^{+0.016}$ & $11.462_{-0.048}^{+0.043}$\\
			$\log M_{\Gamma, \rm s}$ & $10.69_{-0.47}^{+0.51}$ & $11.090_{-0.011}^{+0.010}$ & $11.436_{-0.093}^{+0.044}$ & $11.466_{-0.100}^{+0.092}$ & $11.25_{-0.51}^{+0.13}$ & $11.40_{-0.23}^{+0.17}$\\
			\hline\hline
			\multicolumn{7}{l}{Type: CSMF, Cosmology: lower $S_8$}\\
			\hline
			$\log M_{\star, 0}$ & $10.870_{-0.087}^{+0.090}$ & $10.74_{-0.11}^{+0.10}$ & $11.03_{-0.12}^{+0.21}$ & $11.233_{-0.114}^{+0.093}$ & $10.91_{-0.13}^{+0.13}$ & $10.43_{-0.43}^{+0.33}$\\
			$\log M_{{\rm h}, 1}$ & $11.72_{-0.18}^{+0.20}$ & $11.61_{-0.21}^{+0.21}$ & $12.09_{-0.21}^{+0.23}$ & $12.27_{-0.14}^{+0.20}$ & $11.95_{-0.18}^{+0.20}$ & $11.31_{-0.65}^{+0.42}$\\
			$\gamma_1$ & $3.46_{-0.96}^{+1.00}$ & $3.52_{-1.02}^{+0.98}$ & $3.57_{-1.03}^{+1.00}$ & $3.65_{-0.98}^{+0.90}$ & $3.49_{-0.97}^{+1.00}$ & $3.46_{-0.93}^{+1.00}$\\
			$\gamma_2$ & $0.228_{-0.025}^{+0.023}$ & $0.325_{-0.022}^{+0.020}$ & $0.269_{-0.105}^{+0.031}$ & $0.174_{-0.043}^{+0.049}$ & $0.338_{-0.035}^{+0.033}$ & $0.466_{-0.043}^{+0.037}$\\
			$\sigma_{\log M_\star}$ & $0.1662_{-0.0098}^{+0.0091}$ & $0.1647_{-0.0097}^{+0.0088}$ & $0.1966_{-0.0084}^{+0.0094}$ & $0.2285_{-0.0078}^{+0.0072}$ & $0.1989_{-0.0118}^{+0.0100}$ & $0.200_{-0.035}^{+0.022}$\\
			$b_0$ & $-1.05_{-0.83}^{+0.58}$ & $-1.56_{-0.60}^{+0.49}$ & $-1.56_{-0.67}^{+1.16}$ & $-1.05_{-0.73}^{+0.71}$ & $-1.68_{-0.51}^{+0.53}$ & $-1.62_{-0.57}^{+0.81}$\\
			$b_1$ & $0.69_{-0.50}^{+0.71}$ & $0.53_{-0.37}^{+0.47}$ & $0.62_{-0.42}^{+0.45}$ & $0.47_{-0.34}^{+0.59}$ & $0.60_{-0.40}^{+0.46}$ & $0.64_{-0.40}^{+0.57}$\\
			$b_2$ & $0.03_{-0.16}^{+0.11}$ & $0.138_{-0.094}^{+0.071}$ & $0.148_{-0.201}^{+0.091}$ & $0.198_{-0.125}^{+0.078}$ & $0.141_{-0.123}^{+0.099}$ & $0.09_{-0.22}^{+0.14}$\\
			$\alpha_{\rm s}$ & $-1.09_{-0.92}^{+0.69}$ & $-0.89_{-0.27}^{+0.30}$ & $-0.35_{-1.29}^{+0.25}$ & $-0.88_{-1.02}^{+0.64}$ & $-0.43_{-0.52}^{+0.26}$ & $-0.69_{-1.09}^{+0.44}$\\
			$f_\Gamma$ & $0.915_{-0.091}^{+0.059}$ & $0.952_{-0.062}^{+0.035}$ & $0.943_{-0.063}^{+0.040}$ & $0.925_{-0.071}^{+0.052}$ & $0.861_{-0.096}^{+0.088}$ & $0.60_{-0.19}^{+0.23}$\\
			$\sigma_\Gamma$ & $0.120_{-0.015}^{+0.017}$ & $0.0875_{-0.0070}^{+0.0068}$ & $0.123_{-0.012}^{+0.036}$ & $0.224_{-0.014}^{+0.014}$ & $0.166_{-0.016}^{+0.020}$ & $0.261_{-0.027}^{+0.021}$\\
			$\log M_{\Gamma, \rm c}$ & $11.125_{-0.016}^{+0.015}$ & $11.2619_{-0.0080}^{+0.0083}$ & $11.256_{-0.012}^{+0.011}$ & $11.217_{-0.019}^{+0.018}$ & $11.233_{-0.018}^{+0.017}$ & $11.468_{-0.045}^{+0.040}$\\
			$\log M_{\Gamma, \rm s}$ & $10.56_{-0.38}^{+0.42}$ & $11.089_{-0.013}^{+0.012}$ & $11.04_{-0.11}^{+0.39}$ & $11.440_{-0.112}^{+0.097}$ & $10.84_{-0.58}^{+0.38}$ & $11.24_{-0.22}^{+0.26}$\\
			\hline
		\end{tabular}
		\caption{Posterior predictions for the galaxy-halo parameters in section \ref{sec:discrepancy} and \ref{subsec:cosmology}. In all cases, we show the median and the $16$th to $84$th percentile range.}
		\label{tab:gal_posteriors}
	\end{table*}

    \section{Excess surface density with discrete density tracers}
    \label{sec:new_delta_sigma}
	
	In this section, we describe our algorithm for computing the excess surface density $\Delta\Sigma$ around galaxies in cosmological simulations. One generic problem is that the surface density would, in principle, require a smooth density field, whereas the matter distribution in cosmological simulations is generally given by discrete points. Additionally, when calculating $\Delta\Sigma$ observationally, we stack all source galaxies in the $r_\rmp$ bin defined by the outer edges $r_{\rmp, 1}$ and $r_{\rmp, 2} > r_{\rmp, 1}$. We would like the $\Delta\Sigma$ computed from simulations to take this binning into account. As we will show below, both problems can be solved simultaneously.
	
	The circularly averaged surface density $\Sigma$ caused by a point particle of mass $M_p$ and projected separation $r_\rmp$ from the source galaxy obeys
	\begin{equation}
        \int\limits_{r_{\rmp, 1}}^{r_{\rmp, 2}} \Sigma(r) 2\pi r \rmd r =
        \begin{cases}
            M_p & \text{if } r_{\rmp, 1} \leq r_\rmp < r_{\rmp, 2} \\
            0 & \text{otherwise}
        \end{cases} \quad \, .
	\end{equation}
	We note that $\Delta\Sigma$ is probed by source galaxies that, in projection, lie around the lens galaxy. Assuming that source and lens galaxies are physically uncorrelated, the distribution of source galaxies should be entirely random. Therefore, the average over $r_\rmp$ when stacking source galaxies corresponds to an area average. Coincidentally, the area-averaged surface density caused by the point particle has a simple form,
	\begin{equation}
	    \begin{split}
            \langle \Sigma \rangle &= \frac{\int_{r_{\rmp, 1}}^{r_{\rmp, 2}} \Sigma(r) 2\pi r \rmd r}{\pi (r_{\rmp, 2}^2 - r_{\rmp, 1}^2)}\\
            &=\begin{cases}
                \frac{M_p}{\pi (r_{\rmp, 2}^2 - r_{\rmp, 1}^2)} & \text{if } r_{\rmp, 1} \leq r_\rmp < r_{\rmp, 2} \\
            0 & \text{otherwise}
            \end{cases} \quad \, .
        \end{split}
	\end{equation}
	\label{eq:new_delta_sigma_1}
    Finally, the average of $\bar{\Sigma}(<r_\rmp)$ obeys
	\begin{equation}
	    \begin{split}
            \langle \bar{\Sigma} (<r_\rmp) \rangle &= \frac{\int_{r_{\rmp, 1}}^{r_{\rmp, 2}} \bar{\Sigma} (<r_\rmp) 2\pi r \rmd r}{\pi (r_{\rmp, 2}^2 - r_{\rmp, 1}^2)}\\
            &= \frac{\int_{\mathrm{max} (r_{\rmp, 1}, \mathrm{min} (r_{\rmp, 2}, r_\rmp))}^{r_{\rmp, 2}} \frac{M_p}{\pi r^2} 2\pi r \rmd r}{\pi (r_{\rmp, 2}^2 - r_{\rmp, 1}^2)}\\
            &=\begin{cases}
                \frac{2 M_p \ln (r_{\rmp, 2} / r_{\rmp, 1})}{\pi (r_{\rmp, 2}^2 - r_{\rmp, 1}^2)} & \text{if } r_\rmp < r_{\rmp, 1}\\
                \frac{2 M_p \ln (r_\rmp / r_{\rmp, 1})}{\pi (r_{\rmp, 2}^2 - r_{\rmp, 1}^2)} & \text{if } r_{\rmp, 1} \leq r_\rmp < r_{\rmp, 2} \\
            0 & \text{if } r_\rmp \geq r_{\rmp, 2}
            \end{cases} \quad \, .
        \end{split}
	\end{equation}
	\label{eq:new_delta_sigma_2}
    Thus, the lensing signal induced by a population of point particles around a group of lenses can be computed by summing $\Delta\Sigma = \langle \bar{\Sigma} (<r_\rmp) \rangle - \langle \Sigma \rangle$ over all lens-galaxy pairs and dividing by the total number of lenses.
    
    We note that the {\sc delta\_sigma} function of {\sc halotools} v0.6 works slightly differently. It does not explicitly take into account that the galaxy-galaxy lensing signal is averaged over a range in $r_\rmp$. Coincidentally, $\Sigma$ is still computed in the same way. However, $\bar{\Sigma} (<r_\rmp)$ is computed for $r_{\rmp, 1}$ and $r_{\rmp, 2}$ and then logarithmic interpolation is used at $r = \sqrt{0.5 \, (r_{\rmp, 1}^2 + r_{\rmp, 2}^2)}$. When applied to the galaxy-galaxy lensing signal in MDR1, we find that our algorithm results in a $\sim 3\%$ higher value for $\Delta\Sigma$, particularly on small scales. Due to the $r_\rmp$ averaging described above, our algorithm should be more exact. Additionally, it has the desirable property of being perfectly linear: $\Delta\Sigma$ around two or more lenses is exactly the average of the individual values for $\Delta\Sigma$.
	
	\section{Accuracy of the analytic model}
	\label{sec:analytic_accuracy}
	
	\begin{figure}
        \centering
        \includegraphics[width=\columnwidth]{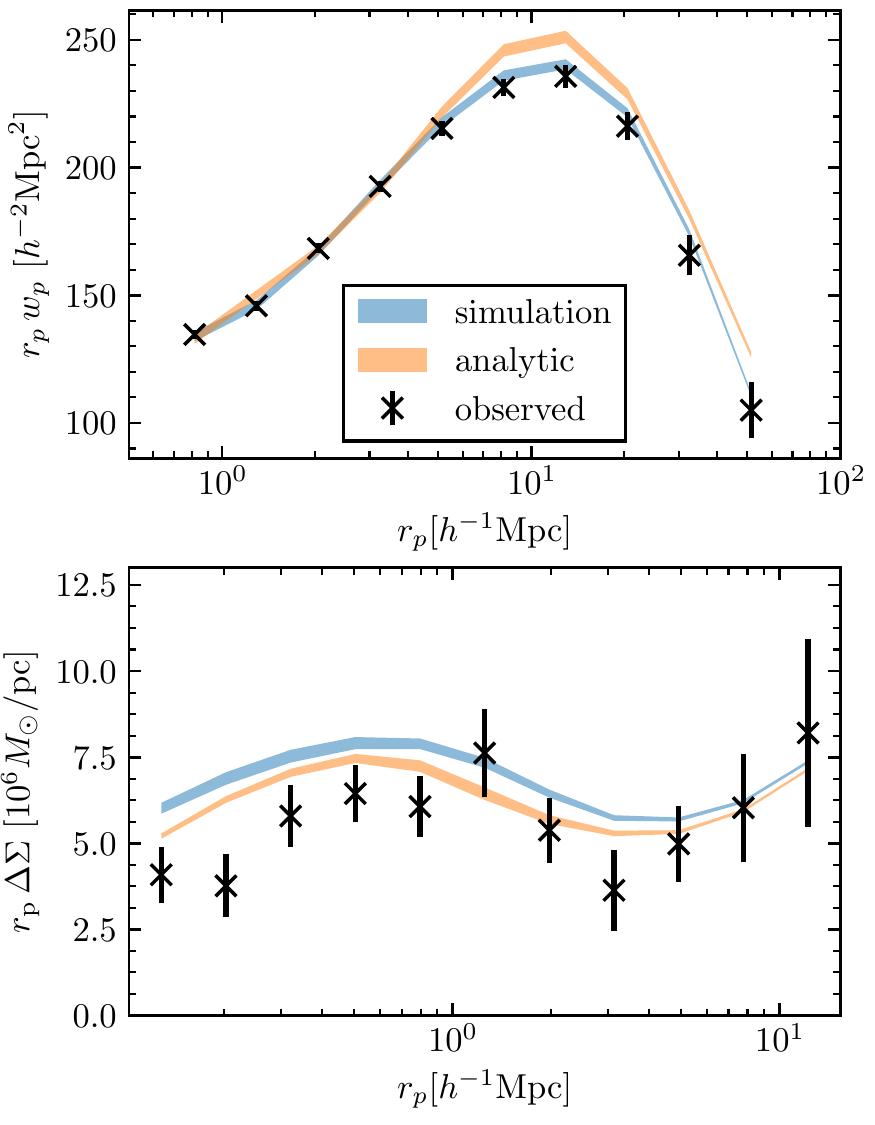}
        \caption{Predictions for the galaxy clustering and galaxy-galaxy lensing for BOSS galaxies in the redshift range $0.5 < z < 0.6$ and masses $\log M_\star \geq 11$. We compare the predictions from the analytic model (orange) to the MDR1 simulation (blue). In both cases, the HOD parameters are tuned to reproduce the galaxy clustering. The analytic model adopts the same cosmological parameters as MDR1.}
		\label{fig:analytic_vs_simulation}
	\end{figure}
	
	Here, we compare the predictions of the analytic model outlined in section \ref{sec:model} to the forecasts of the MDR1 simulation used in section \ref{subsec:galaxy_assembly_bias}. Unfortunately, the analytic model and the MDR1 {\sc ROCKSTAR} halo catalogues do not use the same halo mass definition. Therefore, we cannot directly compare the predictions for $w_\rmp$ and $\Delta\Sigma$ for the same cosmological and HOD parameters. Instead, we choose to compare the predictions for $\Delta\Sigma$ once $w_\rmp$ has been modeled. In principle, this observable should not depend on the halo mass definition because otherwise cosmological inferences would also depend on it.
	
	As in section \ref{subsec:galaxy_assembly_bias}, we choose the $z=0.53$ snapshot from MDR1. For the analytic model, we set the exact same redshift and cosmological parameters as for MDR1. We then use both the analytic model and the simulation to fit the clustering. Fig.~\ref{fig:analytic_vs_simulation} shows the resulting predictions for the galaxy clustering and galaxy-galaxy lensing signal. Overall, the analytic model and the simulation make qualitatively very similar predictions. However, the simulation predicts a lensing signal that is roughly $10\%$ higher on all scales. We find that this difference could be explained by slight, percent-level differences in the halo bias. One should note that our model for the halo bias is the empirical model presented in \cite{Tinker_10}. This empirical model is calibrated with only $2$ simulations with box sizes greater than or equal to $1 (\Gpch)^3$. The bias values measured from these simulations have associated uncertainties stemming from cosmic variance that can explain the small discrepancy seen in Fig.~\ref{fig:analytic_vs_simulation}. Ultimately, the difference could be explained by both inaccuracies in the halo bias model in \cite{Tinker_10} or cosmic variance in MDR1.
	
	We note that the simulation overall predicts an even higher lensing signal at fixed clustering than the analytic model. Therefore, the discrepancies between the predictions and observations described in section \ref{sec:discrepancy} would be even higher for the simulation results.
	
	\label{lastpage}
\end{document}